\documentclass[iop,revtex4,twocolappendix]{emulateapj}

\usepackage{amssymb}
\usepackage[usenames,dvipsnames,svgnames]{xcolor}
\usepackage{lscape} % for landscape figures/tables
\usepackage{numbers}
\usepackage{epstopdf}
\usepackage{amsmath, mathrsfs}
\usepackage{amsfonts}

\definecolor{darkgreen}{rgb}{0,.5,0}

\newcommand{\add}{ } %Indicate new text in a draft
\newfont{\nf}{cmfib8 at 10pt}

\newcommand{\msolar}{\,M$_{\odot}$}

\newcommand{\teff}{$T_{\mathrm{eff}}$}
\newcommand{\logg}{$\log{g}$}
\newcommand{\teq}{$T_{\mathrm{eq}}$}
\newcommand{\Rearth}{$R_{\oplus}$}

\newcommand{\kepler}{{\it Kepler}}
\newcommand{\catrefs}{\citep{Borucki11v1, Borucki11v2, Batalha13, Burke14, Rowe15}}

\newcommand{\nexsci}{NExScI}
\newcommand{\qRange}{Q1-Q16}
\newcommand{\qRangeSmall}{Q1-Q12}
\newcommand{\qRangeSix}{Q1-Q6}

\newcommand{\oldsys}{2412}
\newcommand{\newsys}{3872}
\newcommand{\oldplans}{3136}
\newcommand{\newplans}{4756}
\newcommand{\oldmulti}{480}
\newcommand{\newmulti}{608}
\newcommand{\oldmultiplans}{1204}
\newcommand{\newmultiplans}{1492}

\shorttitle{\kepler\ KOI Catalog VI}
\shortauthors{Mullally et al.}

\begin{document}

\title{Planetary Candidates Observed by \kepler\ VI: Planet Sample from \qRange\ (47 Months)}

\author{F.~Mullally\altaffilmark{1},
Jeffrey L. Coughlin\altaffilmark{1},
Susan E. Thompson\altaffilmark{1},
Jason Rowe\altaffilmark{1},
Christopher Burke\altaffilmark{1},
David W. Latham\altaffilmark{2},
Natalie M. Batalha\altaffilmark{3},
Stephen T. Bryson\altaffilmark{3},
Jessie Christiansen\altaffilmark{4},
Christopher E. Henze\altaffilmark{3},
Aviv Ofir\altaffilmark{5,22},
Billy Quarles\altaffilmark{6},
Avi Shporer\altaffilmark{7,24},
Vincent Van~Eylen\altaffilmark{8},
Christa Van~Laerhoven\altaffilmark{9},
{\add Yash Shah}\altaffilmark{26},
Angie Wolfgang\altaffilmark{10},
W. J. Chaplin\altaffilmark{8,11},
Ji-Wei Xie\altaffilmark{12},
Rachel Akeson\altaffilmark{4},
Vic Argabright\altaffilmark{13},
Eric Bachtell\altaffilmark{13},
Thomas Barclay\altaffilmark{21}
William J. Borucki\altaffilmark{3},
Douglas A. Caldwell\altaffilmark{1},
Jennifer R. Campbell\altaffilmark{14},
Joseph H. Catanzarite\altaffilmark{1},
William D. Cochran\altaffilmark{15},
Riley M. Duren\altaffilmark{16},
Scott~W. Fleming\altaffilmark{17},
Dorothy Fraquelli\altaffilmark{17},
Forrest R. Girouard\altaffilmark{18},
Michael R. Haas\altaffilmark{3},
Krzysztof G. He{\l}miniak\altaffilmark{19},
Steve B. Howell\altaffilmark{3},
Daniel Huber\altaffilmark{1,25},
Kipp Larson\altaffilmark{13},
Thomas N. Gautier III \altaffilmark{16},
Jon Jenkins\altaffilmark{1},
Jie Li\altaffilmark{1},
Jack J. Lissauer\altaffilmark{3},
Scot McArthur\altaffilmark{13},
Chris Miller\altaffilmark{13},
Robert L. Morris\altaffilmark{1},
Anima Patil-Sabale\altaffilmark{14},
Peter Plavchan\altaffilmark{20},
Dustin Putnam\altaffilmark{13},
Elisa V. Quintana\altaffilmark{1},
Solange Ramirez\altaffilmark{4},
V. Silva~Aguirre\altaffilmark{8},
Shawn Seader\altaffilmark{1},
Jeffrey C. Smith\altaffilmark{1},
Jason H. Steffen\altaffilmark{23},
Chris Stewart\altaffilmark{13},
Jeremy Stober\altaffilmark{13},
Martin Still\altaffilmark{21},
Peter Tenenbaum\altaffilmark{1},
John Troeltzsch\altaffilmark{13},
Joseph D. Twicken\altaffilmark{1},
Khadeejah A. Zamudio\altaffilmark{14}}

\altaffiltext{1}{SETI/NASA Ames Research Center, Moffett Field, CA 94035, USA}
\altaffiltext{2}{Harvard-smithsonian Center for Astrophysics, 60 Garden Street, Cambridge, MA 02138 USA​}
\altaffiltext{3}{NASA Ames Research Center, Moffett Field, CA 94035, USA}
\altaffiltext{4}{NASA Exoplanet Science Instititute, California Institute of Technology, Pasadena, CA 91125}
\altaffiltext{5}{Weizmann Institute of Science, Rehovot 76100, Israel}
\altaffiltext{6}{Space Science and Astrobiology Division MS 245-3, NASA Ames Research Center,  Moffett Field, CA 94035 }
\altaffiltext{7}{Jet Propulsion Laboratory, California Institute of Technology, 4800 Oak Grove Drive, Pasadena, CA 91109, USA}
\altaffiltext{8}{Stellar Astrophysics Centre, Department of Physics and Astronomy, Aarhus University, Ny Munkegade 120, DK-8000 Aarhus C, Denmark}
\altaffiltext{9}{1629 E University Blvd, Department of Planetary Sciences, Tucson, AZ,}
\altaffiltext{10}{University of California, Santa Cruz; NSF Graduate Research Fellow}
\altaffiltext{11}{School of Physics and Astronomy, University of Birmingham, Edgbaston, Birmingham B15 2TT, UK}
\altaffiltext{12}{Astronomy and Astrophysics in Ministry of Education, Nanjing University, 210093, China}
\altaffiltext{13}{Ball Aerospace and Technologies Corp., Boulder, CO 80301, USA }
\altaffiltext{14}{Wyle Laboratories, NASA Ames Research Center, Moffett Field, CA 94035, USA}
\altaffiltext{15}{McDonald Observatory and Department of Astronomy, The University of Texas at Austin, Austin, TX 78712}
\altaffiltext{16}{Jet Propulsion Laboratory/California Institute of Technology, Pasadena, CA 91109, USA }
\altaffiltext{17}{Computer Sciences Corporation/Space Telescope Science institute, 3700 San Martin Dr, Baltimore, MD, 21218 USA.}
\altaffiltext{18}{Orbital Sciences Corporation, NASA Ames Research Center, Moffett Field, CA 94035, USA}
\altaffiltext{19}{Subaru Telescope, National Astronomical Observatory of Japan, 650 North Aohoku Place, Hilo, HI 96720, USA }
\altaffiltext{20}{Missouri State University}
\altaffiltext{21}{BAERI/NASA Ames Research Center, Moffett Field, CA 94035, USA}
\altaffiltext{22}{Institut f\"ur Astrophysik, Universit\"at G\"ottingen, Friedrich-Hund-Platz 1, D-37077 G\"ottingen, Germany}
\altaffiltext{23}{Northwestern University Department of Physics and Astronomy
Center for Interdisciplinary Research and Exploration in Astronomy (CIERA)
2145 Sheridan Road, Evanston, IL 60208, USA
}
\altaffiltext{24}{Sagan Fellow}
\altaffiltext{25}{Sydney Institute for Astronomy (SIfA), School of Physics, University of Sydney,
NSW 2006, Australia}
\altaffiltext{26}{{\add Dept. Astronomy, University of California, Berkeley, CA 94720}}
\email{fergal.mullally@nasa.gov}

%``All of science is either physics or stamp collecting''\\
%- E. Rutherford

\begin{abstract}
%Rodgers14 argues that 1.5-1.6Re is a good threshold for rocky/non-rocky threshold
We present the sixth catalog of \kepler\ candidate planets based on nearly 4 years of high precision photometry. This catalog builds on the legacy of previous catalogs released by the \kepler\ project and includes \NumberOfNewKOIs\ new \kepler\ Objects of Interest (KOIs) of which \NumberOfNewPCs\ are planet candidates, and \NumberOfNewSmallPc\ of these candidates have best fit radii $<1.5$\,\Rearth.
This brings the total number of KOIs and planet candidates to \NumberOfKOIsTotal\ and \NumberOfPCsTotal\ respectively.  We suspect that many of these new candidates at the low signal-to-noise limit may be false alarms created by instrumental noise, and discuss our efforts to identify such objects. We re-evaluate all previously published KOIs with orbital periods of $>50$ days to provide a consistently vetted sample that can be used to improve planet occurrence rate calculations. We discuss the performance of our planet detection algorithms, and the consistency of our vetting products. The full catalog is publicly available at the NASA Exoplanet Archive.
\end{abstract}

%%Version number and table of contents are useful for draft versions.
%\section*{Todo}
%\small
%\begin{verbatim}
%$Id: ms.tex 58207 2015-02-03 22:18:59Z fmullall $
%x Add Yash as co-author
%- List of single transit events?
%- Mention list of 440 in abstract
%\end{verbatim}
%\small
%\section*{Table Of Contents}
%\tableofcontents

%\setlength{\parskip}{1.5ex plus0.5ex minus0.2ex}

\section{Introduction}

The NASA \kepler\ mission \citep{Koch10} offers an unprecedented view of the time-domain Universe. Following \kepler's launch in 2009, it obtained nearly uninterrupted observations of a single large field (over 100 square degrees) centered 13.5$^{\circ}$ above the Galactic plane. These long duration, near continuous observations are needed to achieve the mission goal of detecting the transit of Earth size planets in the habitable zone of Sun-like stars.

\kepler\ data is a uniquely valuable observational resource.
The volume of data has encouraged the development of new techniques to detrend time series data \citep{Coughlin12, Roberts13, Ambikasaran14, Handberg14}, and facilitates science not directly related to the core \kepler\ science of finding planets. Highlights include asteroseismology of main-sequence stars and red giants, \citep[e.g.,][]{Chaplin14, Bedding11}, classical pulsators \citep[e.g.,][]{Szabo10}, eclipsing binaries \citep[e.g.,][]{Conroy14}, ages of clusters \citep{Meibom11}, and  active galactic nuclei \citep{Mushotzky11}.

The project has produced several catalogs of planet candidates \catrefs.
These catalogs provide timely updates of new, interesting individual objects suitable for ground-based follow-up, and the necessary data for planet occurrence rate calculations as a function of radius, orbital period and other properties \citep[e.g.,][]{Youdin11, Catanzarite11, Howard12, Morton12, Dressing13, Dong13, Mulders14}.
Independent groups have constructed their own catalogs \citep{Petigura13, Schmitt14, SanchisOjeda14}, and derive their own occurrence rates \citep{ForemanMackey14}. The comparison of these independent efforts will help identify and correct the inevitable insufficiencies  that any approach will suffer for a data set of this size and richness.
The true false positive rates of our catalogs are not known \citep[e.g.,][]{Dressing14, Lissauer14, Rowe14multis,Fressin13, Morton11}, and this must be addressed when deriving occurrence rates.

Observations of the \kepler\ field ended on 2013 May 11 with the failure of the second of four on-board reaction wheels. The spacecraft could no longer maintain the required pointing precision in the original field and was re-purposed for an ecliptic plane mission \citep[K2;][]{Howell14}.

The catalog presented here is the first based on an essentially complete data set on the original \kepler\ field
{\add (one more month of data was collected after work on this catalog was started, but does not noticeably add to the timespan of observations). Future catalogs will include the complete data set, but will focus on finding new planets through improvement in our detection algorithms instead of relying on the longer baseline.}
With 4 years of data, we become sensitive for the first time to transit depths  $\sim 200$\,ppm at periods of 200 days or more -- the parameter space of Earth analogs around Sun-like stars, and an important goal of the mission.

In this article, we report on our efforts to detect and vet new planet candidates with 46 months of data. We also re-vet known candidates from older catalogs with periods $>$50\,days to provide a consistently vetted sample to aid occurrence rate calculations. This period cut eliminates most of the false positives due to eclipsing binaries. We add \NumberOfNewKOIs\ new objects of interest to the previous catalogs, of which  \NumberOfNewPCs\ we deem to be valid planet candidates. We discuss the reliability of these candidates in \S~\ref{human}. In an effort to reduce the subjectivity inherent in the human-dominated false positive identification techniques previously used, we begin to apply objective, rule-based tests to improve the uniformity of our sample, a process that will mature in later catalogs. We make our catalog publicly available at the NASA Exoplanet Archive (see \S~\ref{NEXSCI}). Readers whose primary interest is in making use of this table may wish to skip to \S~\ref{CAVEATS} for a discussion of the features of the catalog they should be aware of.

\section{Input Data}
%Note: This is assuming DR24 hits the MAST before my paper hits the journal
The \qRange\ catalog is based on analysis of 16 quarters of data obtained by the \kepler\ spacecraft from 2009 May 13 to 2013 April 8. A total of \TotalNumberOfObservedStars\ targets were observed, with \TotalNumberOfStarsObservedMoreThanThirteenQuarters\ observed for at least 13 quarters \citep{Tenenbaum13}. The processed lightcurves and calibrated pixel images from which the lightcurves are derived are available at the Mikulksi Archive for Space Telescopes (MAST)\footnote{\url{http://archive.stsci.edu/kepler}}. Note that the lightcurves available at MAST (Data Release 24) come from a newer version of the pipeline than was used in this analysis, but the differences are not substantial. Commonly observed artifacts of the data are described in the Data Characteristics Handbook \citep{Christiansen13dch}, and important features of the quarterly data are described in the Data Release Notes for that quarter \citep{Thompson13drn24, Thompson13drn21}. Both the Characteristics Handbook and the Release Notes can be found on the MAST webpage.

\section{Pipeline Processing}

This data set was reduced and analyzed by version 9.1 of the Science Operations Center pipeline \citep{Jenkins10}. The components of the pipeline are referred to by short 2-3 letter names. Calibrated pixel fluxes were generated by the CAL module \citep{Quintana10}, lightcurves were extracted by the PA module \citep{Twicken10}, and systematic error removal was performed by the PDC module \citep{Stumpe12, Smith12, Stumpe14}. The generation of lightcurves does not significantly distort the transit signals. \citet{Christiansen13} injected simulated transits into \kepler\ data and found that ($< 1$\%) of single transits were suppressed by these three pipeline components, while the signal to noise ratio of the remainder were preserved to better than 0.3\% on average.

Potential transits are identified and characterized by the TPS \citep{Tenenbaum14} and DV \citep{Wu10} modules. These events are known as Threshold Crossing Events, or TCEs. A list of the \NumberOfTces\ TCEs found in this pipeline run is available at the NASA Exoplanet Archive (\S~\ref{NEXSCI}).

While an estimate of the number of transit signals not detected by TPS (the false negative rate) is a topic of active research \citep{Christiansen13}, the fraction of detections not due to transits (the false alarm rate) is high. Of \NumberOfTces\ TCEs found by TPS in this processing, detailed examination determined only \NumberOfFederatedKoisInQSixteen\ were transit-like. This is intentional -- the science cost of a false alarm is very much less than that of a false negative. The core of this article details our efforts to winnow valid planet candidates from the long list of TCEs.

\subsection{Characteristics of the TCE sample \label{tpsProblems}}
Testing of the TPS module by injecting simulated transits suggests that the recovery rate is close to 100\% for TCEs with multiple event statistic (MES) $\gtrsim 16$, but falls to 0 by MES of 6 \citep{Christiansen15}, consistent with the estimate of \citet{Fressin13}. MES is defined in detail in \citet{Jenkins02mes}, but is equivalent to the signal-to-noise ratio of the transit detection in the folded lightcurve. {\add A MES $> 7.1$ is required by the pipeline for detection of a TCE.}
TPS is known to under-detect short period ($\lesssim 10$\,day) planets. Short period planets are often mistaken for coherent stellar variability, which is removed by TPS before searching. Removing this variability increases the yield of longer period, more interesting, planets that would otherwise be missed.

\begin{figure}
    \begin{center}
   \includegraphics[angle=0, scale=.7]{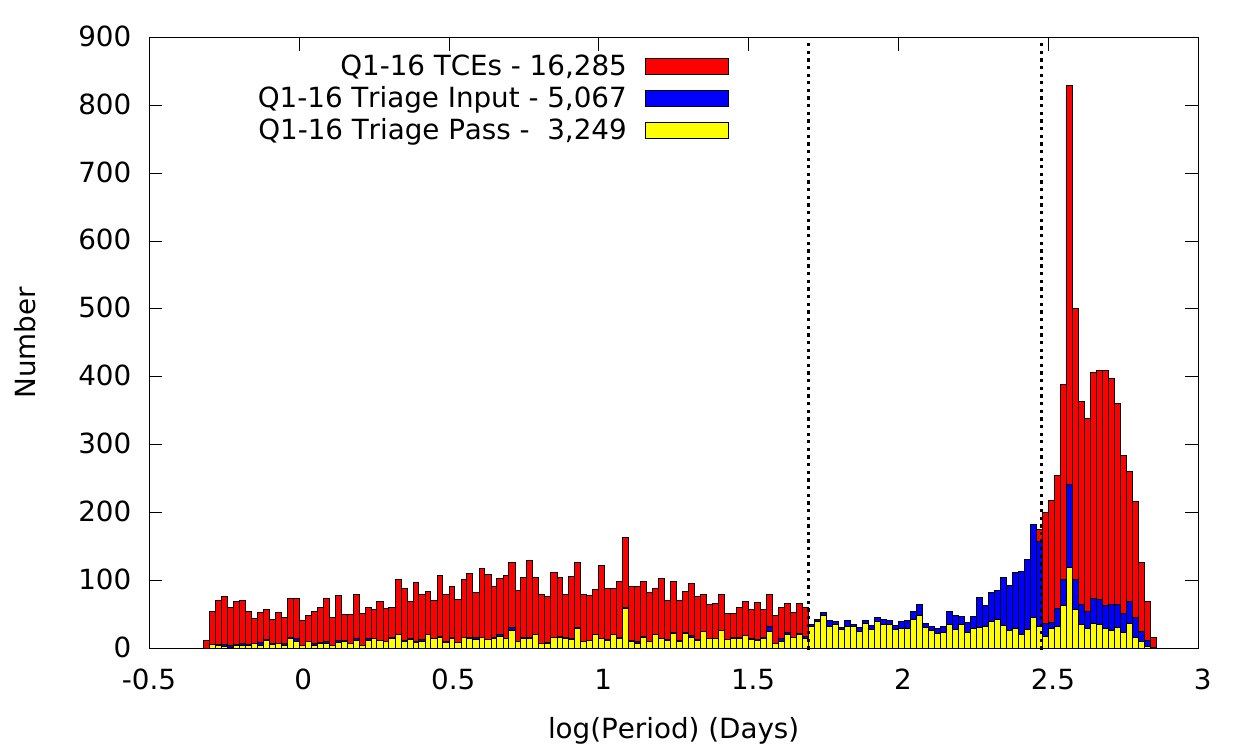}
   \caption{Histogram of number of TCEs found by the pipeline as a function of $\log_{10}$ Period (red histogram). The sharp spike near $\log_{10}$ Period $= 2.6$ is due to rolling band noise, and the broader peak is due to other sources of three-transit false alarms (see \S~\ref{tpsProblems}). The cuts detailed in \S~\ref{jeffCuts} reduce the sample to the blue histogram, while the manual triage process reduces it to the yellow histogram. No cuts were made for TCEs with period of 50-300 days (between the vertical dashed lines), hence the sudden drop in the number of TCEs requiring triage (blue histogram) at 300 days. \label{triageCuts}}
    \end{center}
\end{figure}

Two different kinds of instrumental artifacts dominate the TCE population at long  periods ($\gtrsim 200$\,days).
Edge effects around gaps, stellar flares, and short-timescale systematics can cause transit-like signals in the whitened lightcurve that are detected at low significance. These can combine to produce a significant detection that may not be rejected by the various tests TPS uses to identify false alarms \citep{Seader13, Tenenbaum13}
The probability of finding $n$ equally spaced events of roughly equal significance drops rapidly with increasing $n$ (or decreasing period), so these systematics are predominantly an issue for TCEs with three events, or periods of hundreds of days. In fact, we believe the number of false alarm TCEs with three detected transits is many times greater than the number of planet candidates detected with three transits.
This issue, first noted in \citet{Tenenbaum14}, causes the broad peak in the TCE histogram shown in Figure~\ref{triageCuts}. We discuss our mitigation efforts in \S~\ref{jeffCuts}.

The second source of systematics creates TCEs with periods close to the orbital period of the spacecraft.
The spacecraft rotates around its boresight 4 times per orbit to keep the solar panels pointed at the Sun. Each star is observed by up to 4 different {\sc ccd}s, each with their own noise properties. This causes an abundance of artifacts with periods similar to the orbital period (\KeplerOrbitDays\ days). In particular, a small number of channels suffer from
unstable readout amplifiers that cause background levels to vary in time and {\sc ccd} position \citep{Caldwell10}. These ``rolling bands'' cause small amplitude variations in the background level underneath stars that are flagged by TPS as TCEs.

Finally, poorly corrected systematic signals in the lightcurves can also produce false alarm TCEs, although TPS has considerably improved its ability to identify and eliminate these problems. We discuss our efforts to cull these various false alarms from our catalog in the next two sections.

\section{Triage}
Distilling a population of planet candidates from the list of TCEs is the focus of the work described in this article, and is performed by the Threshold Crossing Event Review Team (TCERT). Following \citet{Burke14}, we take a two-tiered approach. The first tier is to quickly eliminate the bulk of TCEs that can be readily identified as false alarms (defined here as TCEs not produced by the transit of one astrophysical body across the face of another). This tier, called triage, is discussed in this section. Triage proceeds in three steps: Federation, where TCEs associated with known KOIs are found; rejection of false alarms by rule; and visual examination of the remainder. TCEs that pass triage are given a KOI number and pass to the second step, vetting, where we apply further scrutiny to potential false alarms, and identify false positives, or transit events caused by binary stars. Vetting is described in the next section.

\subsection{Federation \label{federation}}
The federation process involves identifying TCEs corresponding to pre-existing KOIs based upon the two predicting similar in-transit cadences.
Periods and epochs that agree within the uncertainties may still predict different sets of in-transit cadences over the 4 years of \kepler\ data.
Instead of simply comparing the ephemerides, we test whether the in-transit cadences predicted by the two ephemerides agree.

For each KOI, and for each observed cadence, we set a boolean value, $y_{i,\rm KOI}$ to true if the midpoint of the transit falls in that cadence. We also calculate the number of transits of the KOI that are predicted to occur during the observations, $N_{\rm KOI}$.

Next, we generate another boolean vector, $y_{i,\rm TCE}$, of transit events for each TCE detected around the same target where the period of the TCE is within a factor of three of the period of the KOI. Unlike the single cadence marking the KOI transit events, this vector is set to true for {\em all} cadences that occur within the transit duration of the mid-transit time as estimated by the TCE ephemeris. We also define $N_{\rm TCE}$, the number of transits of the TCE that occur during the observations.

Finally, we measure the degree of correlation between the KOI and TCE transits,

\begin{equation}
E=\frac{\Sigma y_{i,\rm KOI}    \times y_{i,\rm TCE}}{N_{\rm TCE}}-\frac{\Sigma y_{i,\rm KOI}\times \widetilde{y_{i,\rm TCE}}}{N_{\rm KOI}}\label{eq:fedcorr}
\end{equation}
where the summation is over the observed cadences and
$\widetilde{...}$ indicates the logical NOT operation (i.e., the vector indicating out-of-transit cadences according to the TCE
ephemeris).

The first term of Equation~\ref{eq:fedcorr} measures the degree to
which the KOI transit events align with the TCE transit events, and
the second term penalizes $E$ for having KOI transit events align
with data out-of-transit according to the TCE. The ephemeris correlation
statistic ranges from $-1\leq E\leq 1$, where a value of 1 indicates
that all predicted KOI transit events overlap within a transit
duration of the TCE ephemeris, and $-1$ indicates no correlation.
We require $E>0.8$,
indicating that at least 90\% of the KOI transit events agree with the
TCE, to federate a TCE to a KOI. This high level of federation ensures that
the cadences involved in calculating the vetting metrics in DV
correspond to a similar set of cadences being used for identification
and analysis of the KOI from previous pipeline runs.

Transits from planets in multi-planet systems may not be equally spaced in time due to gravitational perturbation of the orbit. In extreme cases, TPS will not find such events. In less extreme cases, TPS may generate a TCE, but that TCE will fail federation.

\begin{figure*}
     \begin{center}
    \includegraphics[angle=0, scale=.6]{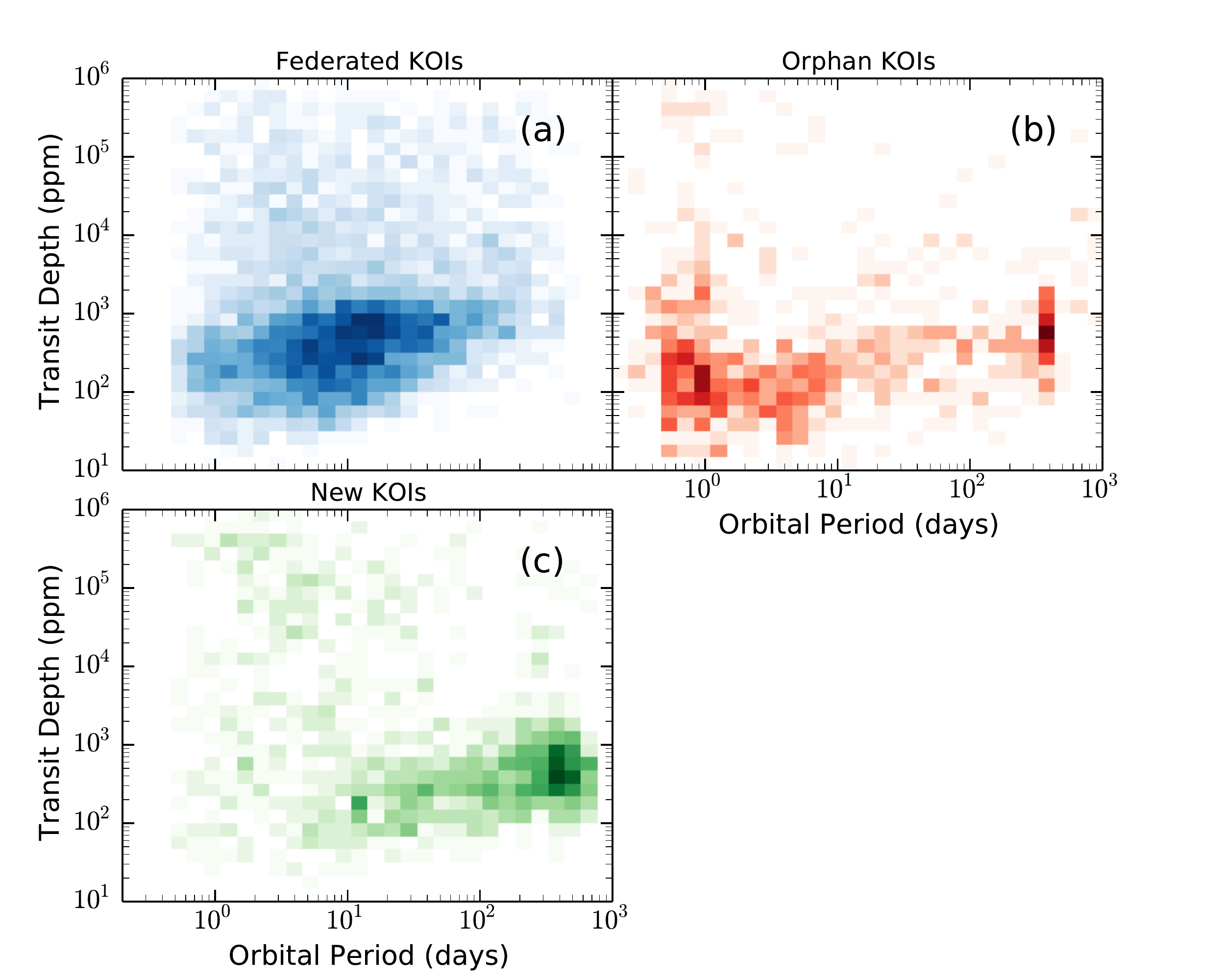}
    \caption{{\bf (a):} Distribution of KOIs previously discovered in earlier catalogs, and recovered in this catalog. We refer to these as ``federated'' KOIs. The bins are equally spaced in the logarithm of orbital period and transit depth. Darker colors indicate a greater density of KOIs in that bin. Both planet candidates and false positives are included in this plot. {\bf (b):} Same as (a), but for previously known KOIs not recovered in \qRange. These KOIs are called orphans. {\bf (c):} KOIs newly discovered in this catalog. For the first time, the KOI population extends to transit depths below 200\,ppm for periods longer than 300 days.\label{recover}}
     \end{center}
 \end{figure*}

\subsubsection{Recoverability of KOIs \label{orphan}}
A good indication of the performance of TPS in a given region of parameter space is its ability to rediscover KOIs found with previous versions of the pipeline and smaller data sets.
In the top left panel of Figure~\ref{recover}, we show the distribution of KOIs discovered in previous catalogs \catrefs , and also found in run of TPS. For clarity, we don't show individual KOIs, but the density of KOIs in equally-spaced logarithmic
bins of transit period and depth. The panel on the right shows the distribution of KOIs listed in previous catalogs that were {\em not} rediscovered. These un-recovered KOIs are known as ``orphans''. The bottom left panel shows the distribution of newly discovered KOIs.

As expected, the new KOIs cluster at longer periods and shallower depths than the federated KOIs. There is also an additional population of deep, short period transits.
Previous catalogs excluded all eclipsing binary (EB) candidates identified by \citet{Prsa11} and \citet{Slawson11}. This (the EB) candidate list has some overlap with planet candidates, so for the \qRange\ pipeline run  we only excluded EB candidates with
a readily distinguishable secondary eclipse. Thus, the \qRange\ search included many transit-like systems that are likely short period EBs with deep primaries but no detectable secondary eclipse.

The orphan KOIs divide into two main clusters, one clustered around periods of 1 day, and another tightly distributed at a period near \KeplerOrbitDays\ days.
The short period cluster is caused by an over-aggressive pruning of false alarms in TPS, which removed some bona-fide short-period transits. This problem will be fixed in a future version of the pipeline \citep{Seader15}.
The fact that these $\sim$ 1 day period KOIs were not recovered should not be used as a reason to doubt their validity.

The long period cluster (at 300-400 days) is dominated by false alarms due to rolling band noise discussed in \S~\ref{tpsProblems}. These artifacts of correlated noise were mis-identified as KOIs  by \citet{Rowe15}, and were not re-detected with the addition of more data.

If a KOI is not recovered in this pipeline run, it should be examined carefully before committing further resources to its study. However, non-recovery is not sufficient reason to label it as a false alarm. A KOI may be missed because of a change in a pipeline algorithm, or a change in the noise properties of the star, reducing the signal to noise of the transit below our threshold for detection. If the KOI is part of a multiple planet system, it may be exhibiting transit timing variations that can make the KOI difficult for TPS to recover.

\subsection{Rejection of False Alarms by Rule\label{jeffCuts} }
Comparing the distribution of TCE periods from the \qRangeSmall\ and \qRange\ results, there is a large increase in the number of long-period TCEs. This increase is far more than would be expected from analyzing an additional four quarters of data. \citet{Tenenbaum14} found the long-period population was dominated by obviously non-astrophysical false alarms.

Applying the cuts suggested in \citet{Tenenbaum14} to the \qRange\ TCEs reduced the number of TCEs to be vetted  from \NumberOfTces\ to \NumberOfTcesAfterCuts . Of particular importance, it reduced the number of TCEs at periods greater than 300 days from 6,073 to 1,243, thus removing a large number of the long-period false alarms. We show the TCE population before and after the cuts in Figure~\ref{triageCuts}. To facilitate the occurrence rate calculations of \citet{Burke15}, we did not apply the cuts to TCEs with periods 50-300 days.

\subsection{Manual inspection}
After federation, and the cuts mentioned above, we are left with \NumberOfTcesAfterCuts\ TCEs. We subject these TCEs to a visual inspection to remove obvious false alarms. A minimum of two people look at a plot of the folded lightcurve contained in the summary report produced by DV (and available at the Exoplanet Archive) and identify the TCE as either a false alarm or a likely transit. In this, and all subsequent steps, we apply the principle of ``innocent until proven guilty''; a TCE is passed as a KOI unless there is evidence beyond all reasonable doubt that it is not. When the two people disagree on the status, a third person resolves the dispute. Transits likely due to EBs are passed at this stage.

\subsubsection{Post Triage}
A TCE that passes triage is only assigned a KOI number if it stands up to some additional scrutiny. Like any detrending technique, the whitening algorithm used by TPS can exaggerate the depth of some small signals, causing unwanted detections. We therefore require that the transit signal maintains its integrity after an alternative detrending algorithm is applied to the PDC light curve. We employ the non-parametric penalized least squares method from \citet{Garcia10} which includes only the out-of-transit points when computing the filter.

Triage-passing TCEs that show evidence for a transit signal in the alternative detrended light curve are elevated to KOI status and passed on to the vetting stage.
For ambiguous cases, we visually examine the individual transits and check that they are unique relative to other nearby events, uncorrupted by systematic effects, and are consistent in depth and duration.
We elevate triage-passing TCEs to KOI status unless a clear problem is identified.

In total, \NumberOfNewKOIs\ new KOI assignments result from this \qRange\ pipeline analysis. Of these \NumberNewSystems\ are the first KOI around their star, and \NewKOIsInMultis\ are new members of multi-KOI systems.

\section{Vetting}
After triage, we subjected the KOIs to a more rigorous vetting to identify  false positives --- targets that show a transit like event which is not due to the transit of a planet around the target star. Vetting comprises three independent steps. We summarize these steps below, but they are similar to those used by \citet{Batalha13,Burke14}, and \citet{Rowe15}. The products we use for vetting, and a detailed manual for their use, are available at the NASA Exoplanet Archive (see \S~\ref{NEXSCI}).

The large number of KOIs makes it difficult to vet every case. Instead we looked at two overlapping populations: newly identified KOIs, and previously known KOIs with periods greater than 50\,days. The first set provides new targets for follow-up and individual study. The second set provides a uniformly vetted set of planet candidates suitable for occurrence rate studies. Where a previously known KOI with period greater than 50\,days is part of a multi-KOI system (i.e., multiple KOIs found around the same star), we vetted every KOI in that system.

\subsection{Flux Vetting}
Flux vetting seeks to further distill out false alarms that survived triage, and to detect eclipsing binaries by the presence of a secondary eclipse. The transits are visually inspected as individual events, and in folded lightcurves on a per quarter basis, per season basis, and across the full timeseries. KOIs that do not look significant compared to the noise in the data are identified and marked as false alarms.

If the primary transit looks real (i.e., statistically significant and not caused by data artifacts) we proceed to search for evidence of a secondary. The folded lightcurve is convolved with the best fit model transit, and the result is inspected for evidence of a second event. Outliers and noisy data can often confuse this metric, so manual inspection is required for this step. Our flux vetting procedure is described in detail in \citet{Rowe15}.

\subsection{Centroid Vetting \label{centroidVetting}}
Centroid vetting looks for evidence that the source of a transit is inconsistent with the location of the target star on the sky. It proceeds by computing difference images, then comparing the flux in- and out-of-transit on a per pixel basis for each quarter. \citet{Bryson13} describes the centroid vetting in considerable detail, and the approach has remained largely unchanged. Transits from resolved sources within the mask of collected pixels can be identified by visual inspection, while transits that occur on stars unresolved from the target are identified by measuring the change in the location of the star's centroid by fitting model pixel response functions \citep[PRF; the \kepler\ PRF is described in][]{Bryson10prf} to the difference images. What is different with this catalog is that we used a machine learning algorithm dubbed the centroid robovetter \citep[to be described in][]{Mullally15} to automate this procedure for a considerable number of KOIs.
The robovetter is a rule-based system that seeks to emulate the behavior of TCERT using the same data products through a series of objective tests.

DV produces per-quarter difference images of the per-pixel flux change during transit. The robovetter first identifies and rejects noisy difference images, then checks the remaining images for evidence that the transit source is associated with a resolved foreground or background source in the observed pixel mask. If the source of the transit is unresolved from the target star then the algorithm searches for shifts in the photometric centroids during transits using the per-quarter fits to the model PRF created by DV.  In each step it implements a simple logic based on the rules built up by TCERT.
Those KOIs where the automated technique reported a difficulty reaching a conclusion (mostly in the low SNR regime) were audited by two human vetters as before.

Extensive testing on earlier catalogs shows that the robovetter agrees with human determinations at the 90\% level. Most disagreement occurs for ephemeris match false positives (see \S~\ref{ephemMatch}). Neither the robovetter nor manual vetting reliably identifies false positives when the source of the transit is outside the pixel mask. When those KOIs are removed, the agreement with TCERT rises to 98\%. As a sanity check on the robovetter's performance, we checked the results of the robovetter by hand for a small number (100) of KOIs, and found agreement at a similar level.

For very high SNR cases, where the simplifying assumptions used in the transit model break down, the DV fit sometimes fails to converge and no difference images are produced.
In these ($\approx$ 80) cases, we plotted the folded per-pixel lightcurves to determine if the pattern of measured transit depth across pixels was consistent with the transit occurring on the target star. Although a much less precise test than the centroid fits, it still identified a small number of false positive KOIs.

\subsection{Ephemeris Matching \label{ephemMatch}}
Recently \citet{Coughlin14}, hereafter referred to as C14, identified over 100 false positive KOIs not found by other techniques by
finding TCEs that share a common period and epoch with an EB or variable star. This ephemeris matching technique proved especially useful in identifying very low signal-to-noise false positives, which were unidentifiable via other techniques.

For this catalog, we perform ephemeris matching using similar methods and techniques of C14. However, instead of comparing KOIs to EBs, we started with the TCE population, thus comparing TCEs to themselves, KOIs, and EBs. Specifically, we used the following catalog sources:

\begin{enumerate}
\item The list of 16,285 \qRange\ TCEs from \citet{Tenenbaum14}.

\item The list of 7,286 KOIs, ranging from KOI 1.01 to 6251.01, available at the NASA Exoplanet Archive as of 2014 June 9. These include the new KOIs reported in this paper, as well as KOIs from previous \kepler\ catalogs.

\item The \kepler\ eclipsing binary catalog list of 2,522 ``true'' EBs found with Kepler data as of 2014 May 14. The compilation of the catalog and derivation of the fit parameters are described in \citet{Kirk14}.
Previous versions of this catalog are described in \citet{Slawson11} and \citet{Prsa11}.

\item J.M. Kreiner’s up-to-date database of ephemerides of ground-based eclipsing binaries as of 2014 May 14. Data compilation and parameter derivation are described in \citet{Kreiner04}.

\item Ground-based eclipsing binaries found via the TrES survey \citep{Devor08}.

\item The General Catalog of Variable Stars \citep[GCVS,][]{Samus09} list of all known ground-based variable stars, published 2014 April. This catalog includes both eclipsing binaries and other periodic variable stars, such as pulsators.

\begin{figure*}
     \begin{center}
    \includegraphics[angle=0, scale=1.4]{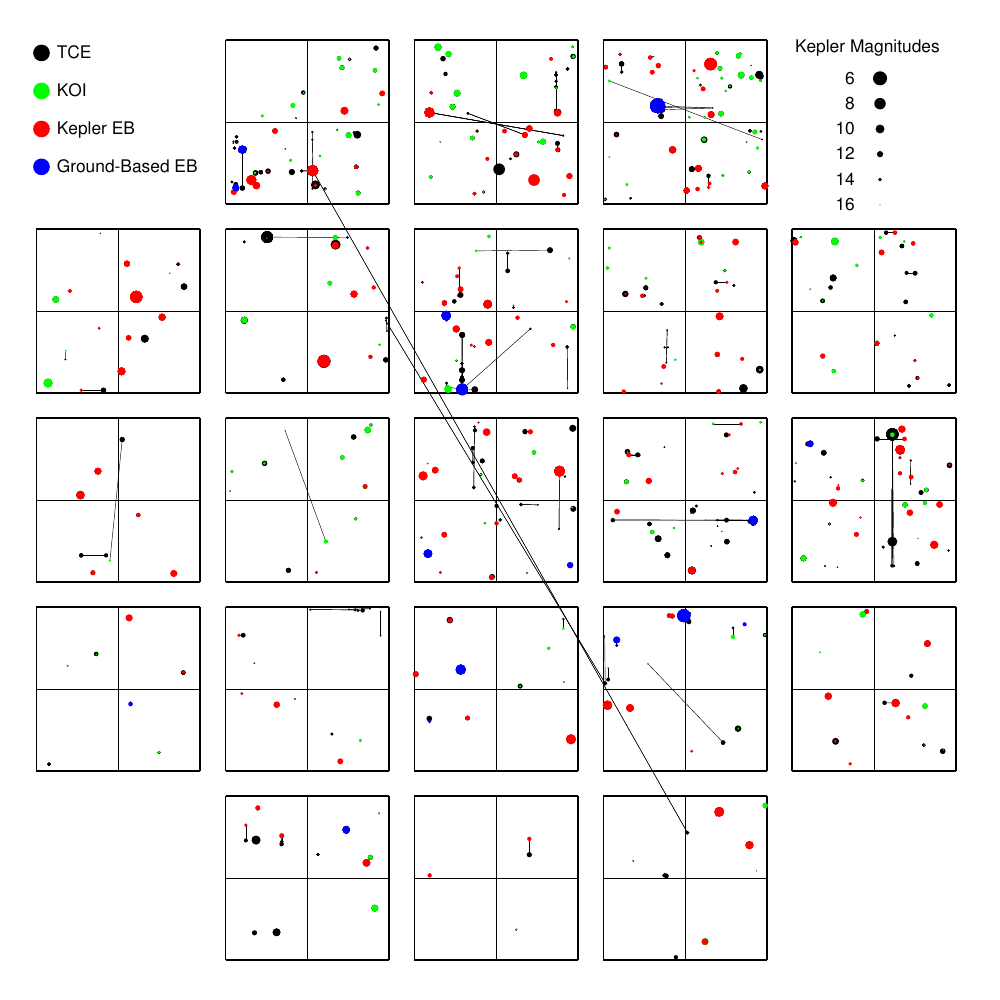}
    \caption{Distribution of ephemeris matches on the focal plane. Symbol size scales with magnitude, while color represent the catalog in which the contaminating source was found. Blue indicates the true transit is from a star in the catalog of \citet{Kreiner04}, and not observed by \kepler. Red circles are stars listed in the \kepler\ Eclipsing Binary catalog, green are KOIs, and black are TCEs. Black lines connect false positive matches with the contaminating parent. In most cases parent and child are so close that the connecting line is invisible.\label{peccd}}
     \end{center}
\end{figure*}

\end{enumerate}

We use the same matching equations as C14 (their Eqns. 1-3). Given the larger number of matches when using the TCE population, and potentially higher number of coincidental matches at high significance, we checked and confirmed that the significance limits for matches used for C14 worked equally well to distinguish between real, statistically significant ephemeris matches, and random coincidences  for our sample.

We ended up with a final list of 960 TCEs with reliable enough matches to designate as false positives. With a longer baseline than C14, some more extreme period ratios are seen, with many at 10:1 or 20:1, and one set of TCEs confirmed to be as high as 45:1. The \qRange\ TCE population contains many false alarms at long periods (see \S~\ref{jeffCuts}).
As a result, we detected ephemeris matches as high as 700:1, (e.g., Kepler ID = 6948098, due to RR Lyrae). However, as these are obvious false alarms at the given TCE period, easily detected as false positives via other means, and time-consuming to separate from coincidence matches at this high of a period ratio, we set our maximum reported period ratio at 50:1, and thus do not record these extreme period ratios ($>$ 50:1) among our list of 960 false positives.

Regarding the mechanism of contamination, the vast majority are still caused by Direct PRF contamination.
Direct contamination from saturated stars can extend out to many hundreds of
arcseconds, due to the large wings of the stellar PRF.
The column anomaly, first reported in C14, caused 74 false positive TCEs, 23 false positives were due to {\sc ccd} cross-talk, and only 2 due to antipodal reflection. In Figure~\ref{peccd}  we plot the location of each false positive TCE and its most likely parent, connected by a solid line. TCEs are represented by black points, KOIs are represented by green points, EBs found by \kepler\ are represented by red points, and EBs discovered from the ground are represented by blue points. The \kepler\ magnitude of each star is shown via a scaled point size. Note that most parent-child pairs are so close together that the line connecting them is not easily visible on the scale of the plot.

Of the 960 ephemeris match TCEs, 625 are KOIs.
The remaining 335 TCEs were either identified as the secondary eclipse of an EB (and not designated as a KOI), or failed triage for some other reason.
Rolling band noise affects nearby stars with the same period and at the same epoch, and thus they can be detected via ephemeris matching. Of the 625 false positive KOIs discovered via ephemeris matching, 138 have been vetted by TCERT in this work, and the remaining 487 are pre-Q16 KOIs with P $<$ 50 days. Of the 138, 17 were classified as planet candidate by TCERT as a result of this current work, and of the 487, 3 were classified as planet candidate as a result of previous work. Thus, ephemeris matching provides false positive dispositions for an additional 487 KOIs in \qRange\ vetting, and corrects the disposition of 20 KOIs from planet candidate to false positive.

\section{Planet Parameters}

\subsection{Stellar Catalog\label{huber}}

The estimate of the radius of a transiting planet depends on the estimate of the stellar radius.
Target selection for \kepler\ depended on the Kepler Input catalog
\citep[KIC,][]{Brown11}. While the KIC can distinguish giant and dwarf stars with reasonable success
\citep[e.g.,][]{Mann12}, the uncertainty in measured stellar surface gravity
propagates into a large uncertainty in inferred planet radius.

\citet{Huber14} describe the latest results of an on-going effort to construct
  an internally consistent catalog of stellar radius drawing from photometry,
spectroscopy and asteroseismology. We use the catalog described in that work
to derive planet parameters. It uses the best information available to estimate
the radius of each individual star by fitting Dartmouth stellar isochrones to observed properties. However, for the majority of the planet-candidate
sample ($\sim 80$\%) the initial estimates of atmospheric properties are
heavily reliant on the KIC. For these stars, the stellar temperatures are updated
using the methodology of \citet{Pinsonneault12} where possible, but the surface
gravity estimates are still based on the KIC narrow-band photometry. Random
uncertainties in KIC derived \logg's have been shown to be $\sim
0.3$\,dex \citep{MolendaZakowicz11, Bruntt12, Huber14}, and several studies have
found evidence that KIC \logg's are systematically overestimated for solar-type
dwarfs \citep{Verner11,Everett13, Farmer13,Bastien14}, meaning the radii are underestimated. These errors will
 introduce biases in the derived planet-candidate radii.

\subsection{Planet Parameters\label{jason}}
We re-measure planet parameters for all our KOIs using the method described by \citet{Rowe14multis}. This provides more realistic uncertainties for our best fit parameters, as well as providing the opportunity to fit the orphans, i.e., KOIs which were not recovered by the pipeline with a longer data set. For previously known KOIs  parameters are unchanged from those reported in \citet{Rowe15}.

We fit all available quarters of data for each KOI, including Q17. The data are cleaned and detrended, using an algorithm that protects the regions known to contain transits. We fit the lightcurve using the transit model introduced in \citet{Seager03} which describes the transit with 5 parameters: period, epoch, impact parameter, the ratio between planet and stellar radius, and stellar density. We use a quadratic limb darkening model from \citet{Claret11} and also include a nuisance parameter to describe any residual mean out-of-transit flux.
 The planet orbit is assumed to be circular.

%MCMC chains run 10^5 times. Source: Rowe et al, draft from 18th Nov 2014 11:07
The uncertainties in our best fit parameters are somewhat correlated, so we use a Markov Chain Monte Carlo (MCMC) approach to improve our error bars, similar to that outlined in \citet{Ford05}. We create 4 Markov chains of $10^5$ fits each and construct our posterior distribution by discarding the first 20\% of each chain. We report the median value at the Exoplanet Archive (\S~\ref{NEXSCI}), and the 1\,$\sigma$ bounds of the distribution as the uncertainty. Transit depth, duration, and planet radius are not directly fit, but computed from the model as described in \citet{Rowe14multis}, properly including the uncertainty in stellar radius.
For many of our false alarm KOIs, the MCMC finds no peak in the posterior distribution for period or epoch, indicating that the fit to the transit is of poor quality. For these KOIs, or others where the MCMC fits fails to give reasonable values, we report only the period, epoch and transit duration of the detection.

\subsubsection{Comparison of MCMC and DV fits \label{jasonVdv}}
\begin{figure}
     \begin{center}
    \includegraphics[angle=0, scale=.4]{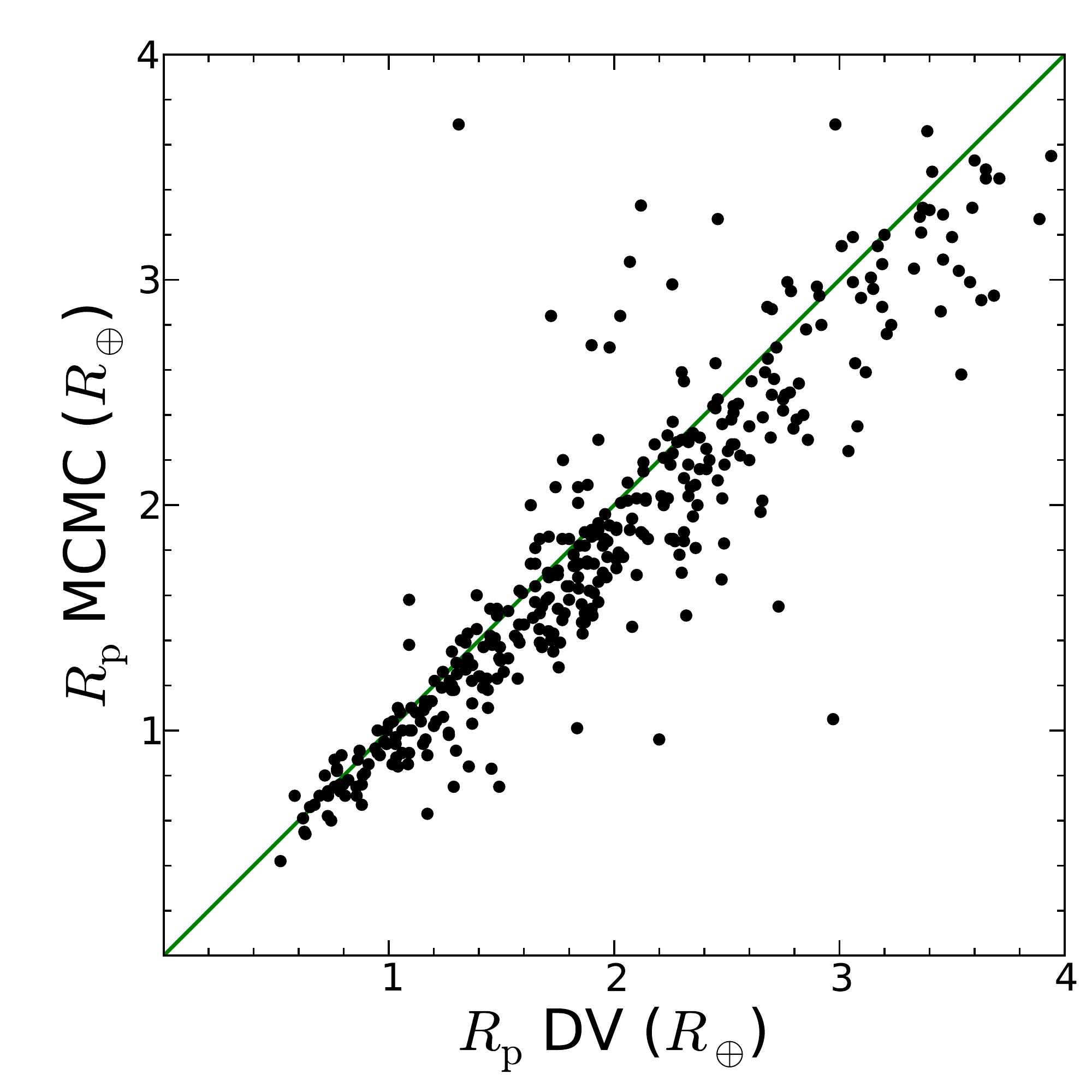}
    \caption{Comparison of measured planet radius from DV and MCMC fits. For clarity, only a small sample of KOIs are shown. The green line shows the one-to-one correspondence. As expected, there is good agreement between the two methods although the MCMC values are systematically smaller than those found by DV. The largest disagreement is typically for KOIs with large fit values for impact parameter.
    \label{rpcomp}}
     \end{center}
\end{figure}

We compare the values from the MCMC fits with those from DV's Marquardt-Levenberg least squares fit \citep{Wu10}. Because the MCMC fits are based on a different detrending, it would be a surprise if their values were identical, but we typically find good agreement between the two approaches (see, Figure~\ref{rpcomp} for an example). However, there are two regions of parameter space where the two approaches tend to disagree.

First, inferred planet radii tend to disagree at short periods. TPS fits and removes coherent sinusoids from the data -- a process called harmonic removal. This allows detection of transits in strongly varying stars, but is known to attenuate the signal of short-period transits, often removing them all together. For short period KOIs ($\lesssim 10$\, days), DV fits should be examined with great care.

Second, both algorithms struggle for v-shaped transits. DV constrains the value of the impact parameter, $b$, to be less than one (corresponding to requiring a small planet to pass within 1 stellar radius of the line of sight). The MCMC fits allow $b$ to float, allowing the radius of the planet to grow arbitrarily large while attempting to fit a grazing transit. We caution that neither approach should be trusted for $b\gtrsim 0.9$.

For lower values of $b$, and for values of $R_{\rm p}/R_{*} < 0.1$ the MCMC estimates of $R_{\rm p}/R_{*}$ are systematically $\approx 7$\% smaller than the values for DV. The source of this difference is under investigation. This bias propagates to the computed values of transit depth and planet radius. Note that the difference between DV and MCMC fits is less than the typical 1\,$\sigma$ MCMC uncertainty in the parameter.

\begin{figure}[t]
     \begin{center}
    \includegraphics[angle=0, scale=.4]{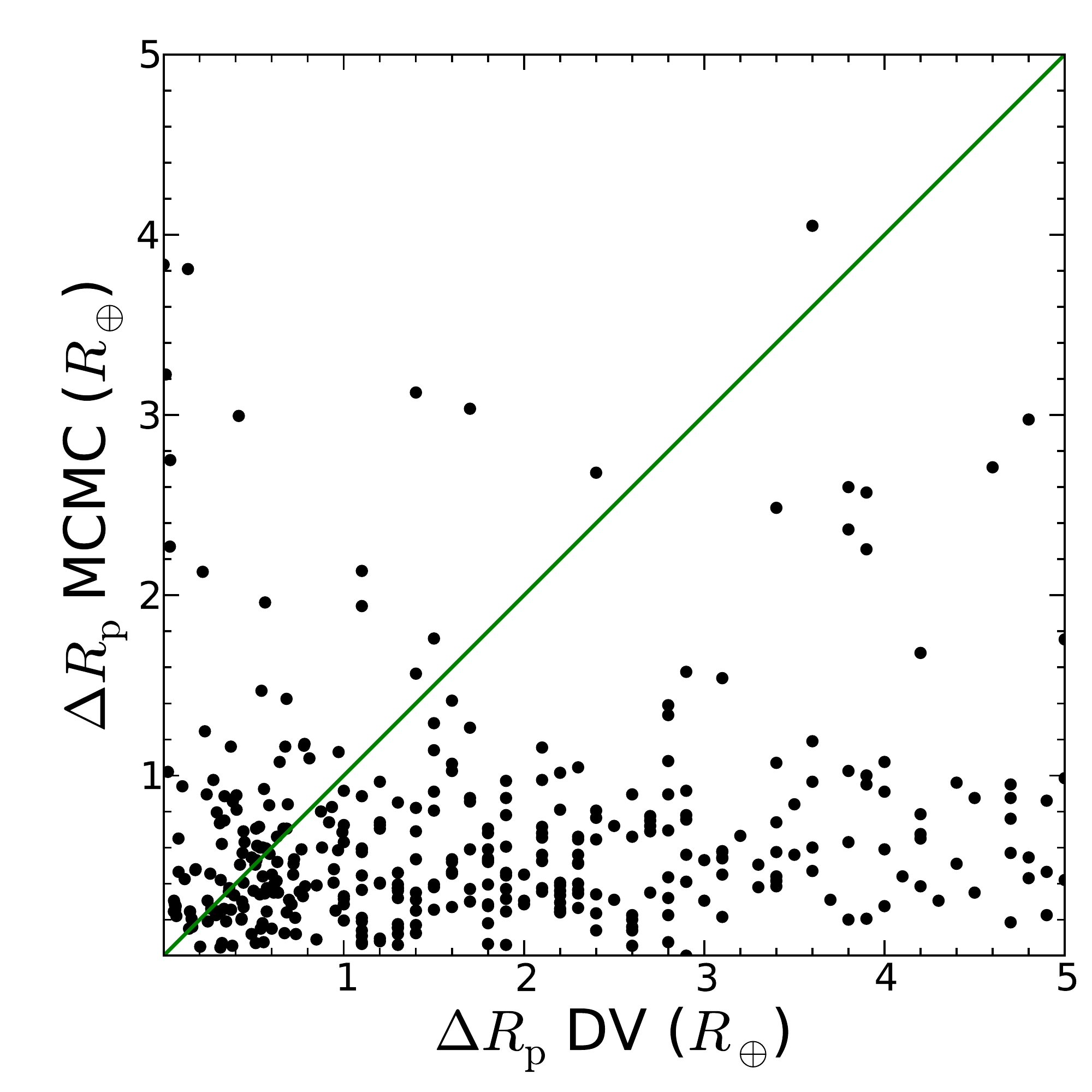}
    \caption{Comparison of the reported uncertainty in planet radius from DV verus the MCMC fits. The green line shows the one-to-one correspondence. The uncertainties reported by DV are systematically larger than from MCMC fits. See text for details.
    \label{rpunc}}
     \end{center}
\end{figure}

The chief advantage of the MCMC approach is the improved uncertainty estimates in the presence of strong covariance between the fit parameters. DV and MCMC uncertainties are in good agreement for parameters that do not covary strongly such as period, epoch, and depth. However, the DV estimate of the uncertainty in $R_{\rm p}/R_{*}$ (which is computed independent of transit depth uncertainty) typically does not agree with the MCMC value.
%These words taken from KSOC-4271. See comment by Burke, 2014-11-18
$R_{\rm p}/R_{*}$ is strongly correlated with $b$, and the local linearized uncertainty
estimate employed by DV becomes ill-conditioned \citep[see Table 2 of][]{Carter08}, resulting in an overestimate of the uncertainty of $b$.  This overestimate in $\Delta b$ propagates into an overestimate in the planet radius uncertainty in DV results (Figure~\ref{rpunc}). {\add The MCMC planet radius uncertainty is preferred to the DV value reported in the TCE tables at the archive (see \S~\ref{NEXSCI}) for this pipeline version.}

\section{Caveats\label{caveats}}
While we make every effort to produce a consistent, well-vetted catalog of planet candidates, a catalog of this size inevitably contains unexpected features. In this section we list the more important caveats to be aware of.

\begin{itemize}
\item The disposition process uses a philosophy of ``innocent until proven guilty''. KOIs are marked as planet candidates if there is insufficient or inconclusive evidence for failure. We therefore expect the set of false positives to be quite pure, but the set of planet candidates will be contaminated with a set of possible false positives, or objects for which few tests were possible. The catalog reliability, or the fraction of false positives lurking in the planet candidate set, is an active area of research.

\item The cuts described in \S~\ref{jeffCuts} eliminated a small number of viable candidates from the sample of TCEs to be triaged. Based on the properties of the known KOIs in previous catalogs, we estimate $\approx$1\,\% of the TCEs removed by this cut would have passed human triage and vetting had they been subjected to it.

\item Most new candidates reported in this catalog were detected with a MES $<$ 10. Easier to detect transits have already been found with fewer data. Unfortunately,  many of the metrics developed by TCERT begin to lose their diagnostic power in this low signal limit, with a corresponding impact on catalog reliability.

\item The false alarm population is dominated by TCEs with three detected transits and a detected MES $<8$. Some tens of these objects survive the various steps of triage and vetting and are labelled as candidates. While there are undoubtedly valid planets in this regime, random chance suggests a significant fraction of the large false alarm population is sufficiently transit-like to survive our tests. We discuss the implications of this population for catalog reliability in \S~\ref{human}

\item The SNR reported by the MCMC parameter fits in \S~\ref{jason} is correlated with the detection MES, but with a large variance. {\add This variance is probably due to the different detrending techniques employed by the pipeline and the MCMC fits.} Many objects detected above the imposed threshold of MES $> 7.1$ have smaller reported SNR values. While a low SNR is a cause for concern for a KOI, it is not sufficient reason to label the transit as a false alarm.

\item The best fit planet radius is not used to identify false positives. Although objects with radii $\gtrsim 200$\,\Rearth are probably too large to be planets for all but the youngest systems \citep{Burrow01}, our estimate of planet radius depends on our classification of the host star. In cases where the host star is a late-type dwarf misclassified as an evolved giant, the planet radius will be dramatically over-estimated.
Given the interest in M stars as planet hosts, and the fact that occasional misclassifications of cool giants and dwarfs cannot be excluded \citep{Brown11, Mann12}, we do not fail such objects based on their inferred radius alone.
%Brown11 See Fig 9 for poor dwarf/giant separation.

\item Our planet parameter fits assume the transiting body is small and non-luminous. When these assumptions are not true, our fits are not reliable.

\item KOIs where the ingress and egress times are a significant fraction of the total transit duration are most likely false positives. These v-shaped transits can be created by EBs for a wide range of inclinations, but only by grazing transits in the case of planets. However, a v-shaped transit by itself is not conclusive evidence of a false positive and these KOIs are marked as planet candidates.
In Table~\ref{newKoiTable} we provide a v-shape metric to identify such cases. The metric is defined as $V = 1- b - R_{\rm p}/R_{*}$, where $b$ is the impact parameter, and $R_{\rm p}/R_{*}$ is the ratio of planet and star radii. KOIs with $V \lesssim 0$ are more likely to be caused by EBs.

%KSOC-3922
\item TPS occasionally identifies EBs at half their true period; in these cases, their false-positive nature can be identified by a statistically significant difference in transit depth between the odd and even numbered transits.
The results of this ``odd-even'' test are available for each KOI in the DV generated reports available at the exoplanet archive (\S~\ref{NEXSCI}).
However, the fits are seeded with the best fit to all transits, which can be a local minimum in the potential space. The depth difference is therefore sometimes underestimated, which may cause a small number of false positives to slip into the planet candidate category. This unintentional bias will be removed in the next version of DV.

\item Users should not rely on the odd-even test for periods $>$90 days.
\citet{vanEylen13} discusses the various reasons why measured transit depths change with {\sc ccd}. In the presence of flux contamination from a brighter star, the differences can be a significant fraction of the transit depth itself.

\item Transits and other variability on bright stars can contaminate targets many tens of arcseconds away \citep[][and \S~\ref{ephemMatch}]{Coughlin14}. The absence of the core of the stellar image in the observed pixel mask of the contaminated star means the centroid tests fail to identify such cases as false positives. In cases where the true source is unobserved, and the false positive signal is only detected on a single target star, ephemeris matching can't identify the target as a false positive. Estimating the number of false positives due to this effect is an ongoing effort.

\item The uncertainty in planet radius is driven by uncertainty in the stellar radius, which can be surprisingly large. For a 1\msolar\ star, an uncertainty in \logg\ of $\pm$0.15 dex, achievable with high-resolution spectroscopy
\citep[e.g.,][]{Valenti05, Ghezzi10},
 translates into an uncertainty in stellar radius of $15-20$\%.
 For stars where the best estimate of the radius comes from the KIC, the error in \logg\ can be much larger.
With asteroseismology, the stellar radius is measurable to within a few percent \citep[e.g.,][]{ChristensenDalsgaard10, Chaplin13, Gilliland13}, and uncertainty in relative planet radius due to stellar activity can become significant \citep{Czesla09}.

%R = sqrt(GM/g) = sqrt(GM)* 10^ax, where a=-1/2, x = logg
%dR/dx = R ln(10)*a
%     = -1.15 R
%=> (\Delta R)/ R = 1.15 \Delta \logg

\item Where a target star is unclassified in the input catalog, the parameter values default to solar: \teff\ = 5780\,K, \logg = 4.438 and stellar radius equal to 1\,$R_{\odot}$. {\add There are 96 KOIs with no KIC parameters, of which 13 are labeled as candidate, of which 7 are newly reported in this paper.}

\item The reliability of the measured planet radius is reduced in crowded fields.
Contaminating light from nearby stars in the optimal aperture reduces the observed transit depth. We attempt to correct for this by estimating the contaminating flux using the method described by \citet{Bryson10tad}. These estimates are known to have issues, especially for brighter stars. Problems with the crowding calculation cause the measured transit depth to change as a function of quarter.

\item Our planet parameters are computed assuming the parent star is single. If the star is a member of a binary system, our fits will tend to systematically underestimate the planet radius.

\item Our planet parameters are computed assuming the planetary orbit is circular. The calculated uncertainties are therefore biased for cases where the orbital eccentricity is non-zero.

\item Most transits are equally spaced in time. In multi-planet systems, gravitational interaction can perturb planet orbits leading to irregularly spaced transits. Such systems are said to exhibit transit timing variations (TTVs) and can be used to confirm the planetary nature of a KOI. The TPS algorithm assumes equally spaced transits and fails to find the most extreme TTV cases. The QATS algorithm \citep{Carter13} is better tuned to find such systems, and \citet{Mazeh13} provides a recent catalog of such systems detected with \kepler. Other complicated systems such as circumbinary planets \citep[e.g.,][]{Doyle11} are typically also not detected.

\item The pipeline requires at least 3 detected transits to claim a detection. Long period planets with fewer than 3 transits are not detected regardless of their SNR. \cite{Kipping14} reports the detection of one such event previously identified by eye as a single transit event in \citet{Batalha13}.

\item Our vetting procedures ultimately rely on human judgment. Despite every care, it is still possible for some errors to slip through.
\end{itemize}

\section{Understanding the Tables at the NASA Exoplanet Archive \label{nexsci}}
The best fit stellar and planetary parameters for all known KOIs are hosted at the NASA Exoplanet Archive \citep{Akeson13}. We show a summary for the KOIs vetted in this paper in Tables~\ref{newKoiTable}~\&~\ref{cjbTable}. The full tables can be accessed through a browser\footnote{Use the URL \url{http://exoplanetarchive.ipac.caltech.edu}} or through the programmers' interface\footnote{\url{http://exoplanetarchive.ipac.caltech.edu/docs/ \newline program\_interfaces.html}}. The table design at the archive reflects our goal of supporting analysis of multiple pipeline runs while serving two distinct community needs: those of follow-up observers \citep[e.g.,][]{Gautier10} who need regular updates of new candidates, and researchers interested in population studies, who need a stable set of KOIs and best parameters from which to do science.

For each pipeline run we post a table of TCEs (from \qRangeSmall\ onward) and a table of KOIs (from \qRangeSix\ onward). The labels given to each catalog, and the appropriate citations are given in Table~\ref{nexsciLabels}. The table corresponding to this paper is labeled Q1-Q16.
The TCE tables contain every event recorded by the pipeline, and the preliminary planet parameters estimated by DV.
The ephemerides, fitted planetary parameters, stellar parameters, and diagnostics  are available for every TCE where that information was measured.
The TCE tables are static and are never updated.

\begin{deluxetable}{lcl}
\tablewidth{250pt}
%\tabletypesize{\tiny}
\tablecaption{TCE and KOI tables at the Exoplanet Archive}
%$Id: nexscitab.tex 58202 2015-02-03 22:10:51Z fmullall $
%$URL: svn+ssh://murzim/repo/so/trunk/vetting/q1q16/paper/nexscitab.tex $

\tablehead{
    \colhead{Table}&
    \colhead{TCE Citation} &
    \colhead{KOI Citation}
}
\startdata
Q1-Q6  & --- & \citet{Batalha13}  \\
Q1-Q8  & --- & \citet{Burke14}  \\
Q1-Q12 & \citet{Tenenbaum13} & \citet{Rowe15}  \\
Q1-Q16 & \citet{Tenenbaum14} & This work  \\
Cumulative & --- & ---
\enddata
\tablecomments{Tables at the Exoplanet Archive and their corresponding citations.
TCE tables for the first two catalogs were not published.
The cumulative table combines results from all other catalogs.\label{nexsciLabels}}
\end{deluxetable}

For each pipeline run, the  KOI table is initially populated by every new (i.e., triage passing) and federated KOI.  The column ``Disposition Using Kepler Data'' is initially set to ``not dispositioned'' to indicate that the KOI has not yet been vetted. As vetting proceeds, this column is filled in with the phrase ``candidate'' or ``false positive'', where false positive also encompasses false alarms (i.e., KOIs that aren't due to the transit of one astrophysical body across another). Dispositions are subject to change as we perform quality assurance on our work.
The diagnostic reports used by TCERT for triage and vetting are accessible from the TCE table and KOI summary pages. A companion guide to help interpret these reports is also available\footnote{\url{http://exoplanetarchive.ipac.caltech.edu/docs/ \newline TCERTCompanion\_q1\_q16.pdf}}.

While changes are possible to the KOI table it is labeled as ``Active''. When work on the catalog is complete, we lock the table and change the label to ``Done''. No further changes will be made to a locked table.

Starting with \qRangeSmall\, we also include 4 flag columns to indicate the reasons a KOI was marked as a false positive. More than one flag can be set simultaneously. In rare cases, a KOI may fail for reasons other than those indicated by the flags, in which case no flag is raised. The flags indicate if a KOI was determined to be:

\begin{enumerate}
\item ``Non-transit like'': A KOI whose light curve is not consistent with that of
a transiting planet. This includes, but is not limited to, instrumental
artifacts, non-eclipsing variable stars \citep[e.g., heartbeat stars, ][]{Thompson12}, and spurious detections.
\item ``Significant Secondary'': A KOI that is observed to have a significant
secondary event, meaning that the transit event is most likely caused
by an EB.
\item ``Centroid offset'': The source of the transit was on a nearby star, not the target KOI
\item ``Ephemeris Match Indicates Contamination'': The KOI shares the same period and epoch as another system and is judged to be a false positive as described in \S~\ref{ephemMatch}.
\end{enumerate}

Planet parameters are initially populated with the best fit values from DV and are replaced with parameters from the MCMC analysis (\S~\ref{jason}) when available. If the MCMC fit fails because the KOI is a false alarm, only the period, epoch, and duration of the detected transit are reported.
MCMC is applied to all KOIs, whether they were found by the pipeline run or not. KOIs shared with the \qRangeSmall\ table have identical fit parameters because, in both cases, 17 quarters of data were available at the time the MCMC fits were performed.
The original fit to the KOI by DV can be found in the corresponding TCE Table by matching the KOI's Kepler ID and planet number.

Since no table is guaranteed to list all known KOIs at any given time, the exoplanet archive also hosts a cumulative KOI table.  This table is generated automatically by the archive and presents the most recent, reliable information available from the individual KOI tables, according to priority lists indicated on the website.  As such, the cumulative table provides the best, most recent, planetary parameters and dispositions on all known KOIs but the parameters and dispositions are currently based on in-homogeneous data sets.  Depending on which KOI you are interested in, the planetary fits and the dispositions are based on a different amount of data and a different version of the pipeline.

\begin{figure*}[!htb]
     \begin{center}
    \includegraphics[angle=0, scale=.45]{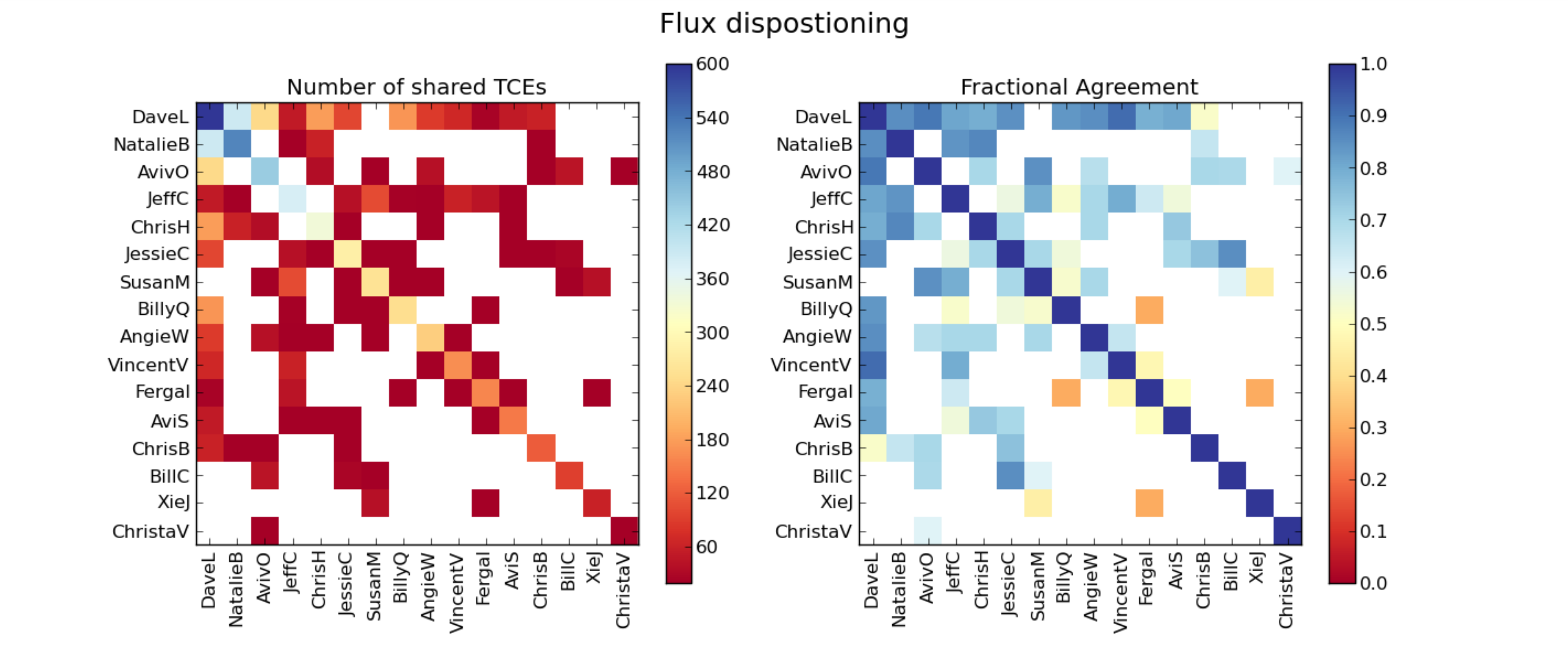}
    \caption{{\bf Left:} Number of KOIs vetted by any given pair of vetters. Blue squares indicate a high overlap in the number of vetted KOIs, red means a low overlap, and white means that fewer than 20 KOIs were shared by a pair of vetters. For clarity, the color scale is pinned at 600 KOIs; the most prolific vetter, DaveL looked at over 1400 KOIs.
    {\bf Right:} Fractional agreement between any two pairs of vetters. There are 3 options (candidate, false positive or ambiguous), so at worst we expect $\sim$30\% agreement. The actual agreement rate is far higher, indicating that our vetters are acting in a broadly consistent manner. The worst correlations (yellow squares) correspond to vetters with small overlap, {\add and indicate a small number of disagreements.}
     \label{fluxvettercorr}}
     \end{center}
\end{figure*}

\begin{figure*}
     \begin{center}
    \includegraphics[angle=0, scale=.45]{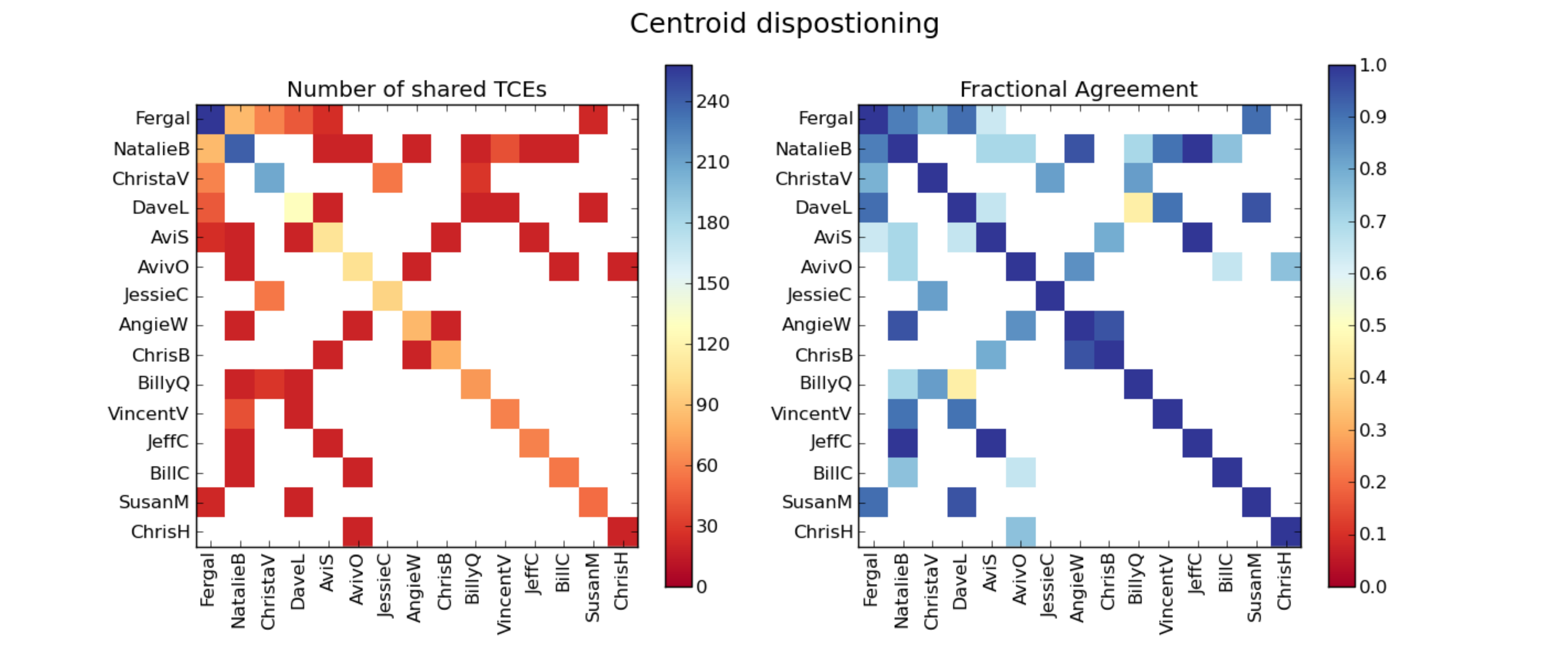}
    \caption{Same as Figure~\ref{fluxvettercorr} but for centroid vetting. Use of the robovetter meant that humans analyzed centroid diagnostics for a smaller set of KOIs.
    \label{centroidvettercorr}}
     \end{center}
\end{figure*}

\section{Discussion}

\subsection{Catalog Reliability\label{human}}
Although we have begun to replace manual inspection with objective tests for triage and vetting, we still rely heavily on human intervention in making dispositions. Our methodology has improved considerably since \citet{Borucki11v1}, but any process involving humans is necessarily subjective, with different people focusing on different features, and the performance of individuals varying with time and circumstance. Every KOI is looked at by at least two different people for flux vetting, as is every KOI the robovetter marked as needing further analysis for centroid vetting. Each person marks a KOI as a planet candidate, a false positive, or as an ambiguous case needing a third opinion.

In Figures~\ref{fluxvettercorr} and \ref{centroidvettercorr} we show the numbers of KOIs shared by any two people and the fractional agreement in initial disposition. The broad agreement indicates that our vetters are acting in a broadly consistent manner, and the disposition of a KOI does not depend strongly on who looked at it. Where two judgments disagreed, or one marked the KOI as needing further analysis, additional people examined the KOI to make the final decision. This ensures the difficult cases get extra scrutiny and considerably improves the consistency of the catalog.

An additional check on our consistency is the 941 objects vetted in this catalog that were also vetted in previous catalogs. Of these, we found 49 KOIs whose dispositions disagreed between this and the previous catalog. Of these, 20 were planet candidates in \qRangeSmall\ that were vetted as false positives in \qRange, and 29 were false positives that were vetted as planet candidates. We re-examined each of these in more detail and confirmed the \qRange\ TCERT dispositions were correct for 36 of them, but the remaining 13 should be overturned to agree with the pre-existing dispositions. This consistency check suggests that our catalogs are self-consistent at the $\sim$ 98\% level.
However, consistent vetting is not the same thing as correct vetting for a given KOI, as alluded to by the caveats listed in \S~\ref{CAVEATS}.

\begin{figure}[thb]
     \begin{center}
    \includegraphics[angle=0, scale=.35]{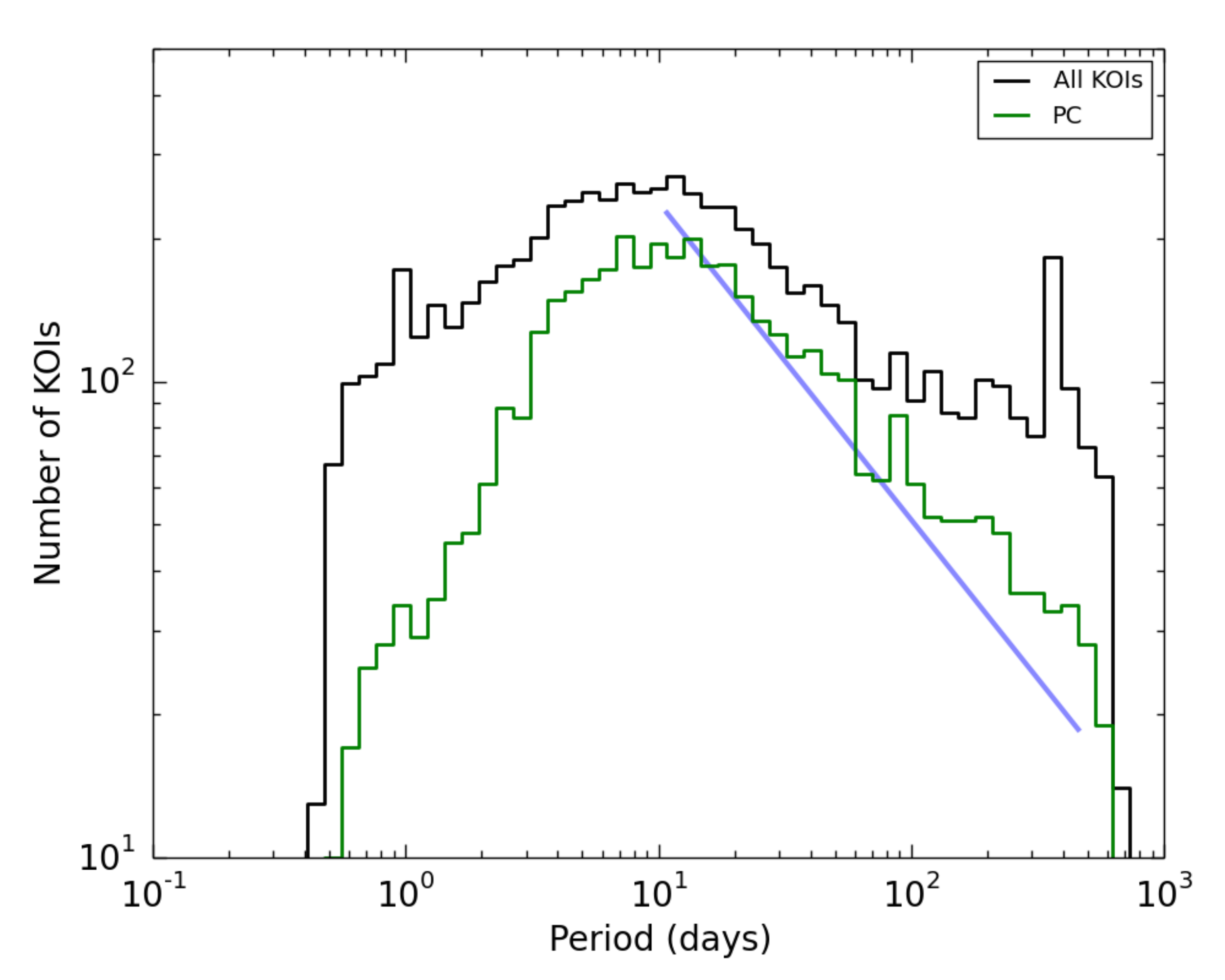}
    \caption{Period distribution of all KOIs found by \kepler\ (upper black line), and all candidates (lower green line, labeled PC). The peak in the KOI distribution near \KeplerOrbitDays\,days due to rolling band noise is eliminated in the candidate sample. The straight blue line shows the expected slope of the distribution if detection rates are driven by the geometric probability of transit. Both the KOI and PC populations show more KOIs at periods $\gtrsim 100$\,days which is likely caused by the population of false alarms that survive triage and vetting respectively. \label{periodDistrib}}
     \end{center}
 \end{figure}

In Figure~\ref{periodDistrib} we show the cumulative distribution in $\log_{10} P$  of all KOIs found in the \kepler\ catalog papers (where $P$ is the orbital period of the KOI). The upper black histogram shows the KOI distribution, while the lower green histogram shows only those KOIs vetted as planet candidates.
The false positive fraction, i.e., the fraction of KOIs not labeled as candidates, is large at both short ($<$ 10\,day) and long ($>$ 100\,day) periods. The short period false positive population is mostly composed of eclipsing binaries, as well as a few variable stars that slipped through triage. At long periods, the KOIs that did not become candidates are mostly instrumental false alarms. It is notable that the peak in the KOI distribution near 372\,days is not reproduced in the candidate sample, providing confidence that TCERT was effective at identifying and rejecting these rolling band false alarms.

For reference, we overplot a power law function proportional to $P^{-2/3}$ (straight blue line). If we naively assume that planets are uniformly distributed in period, and the planet detection rate was driven by the geometric probability of transit, we would expect the planet candidate distribution to follow this trend. (\citet{Pepper03} provides a more detailed prediction of the expected return from a planet survey.) The candidate distribution does follow the trend from 10-100\,days then turns upwards slightly. This may be because planets are more common at $P>100$\,days, but a more prosaic explanation is that the upturn is caused by an excess of false alarm KOIs that pass through vetting in this period range. In previous catalogs we were able to fit our KOIs with more data than was available when they were originally detected, and we used that fact to identify false alarms. The loss of the reaction wheels means that approach is no longer available.

\begin{figure}[!thb]
     \begin{center}
    \includegraphics[angle=0, scale=.35]{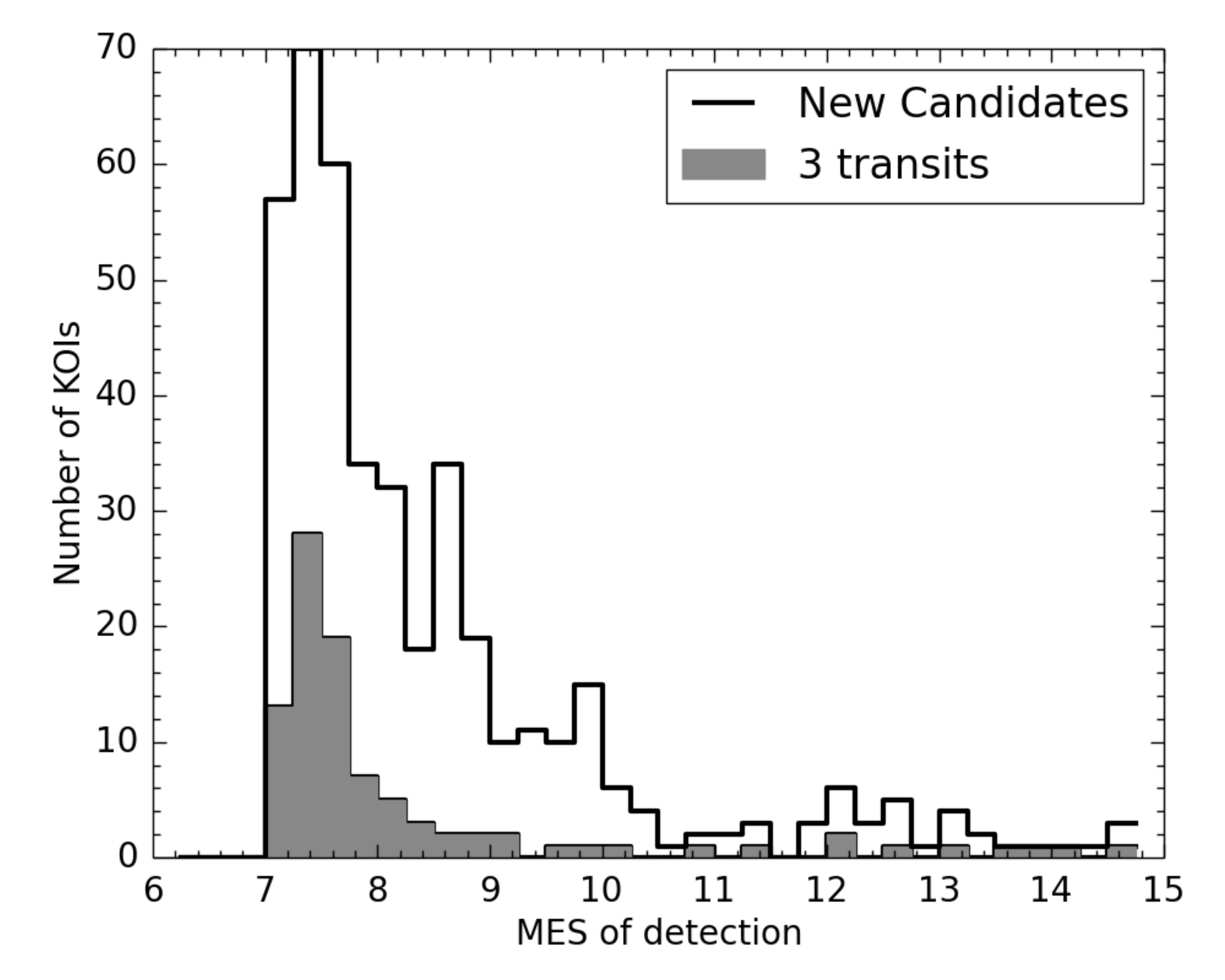}
    \caption{Distribution of the detection statistic (MES) for all new planet candidates listed in Table~\ref{newKoiTable} (white histogram), and the subset of new candidates with only three detected transits (grey). TCEs with three detected transits and low MES constitute the majority of the false alarms.
    At MES $<8$ there are 219 candidates, and 75 have only three detected transits, suggesting the reliability in this low SNR limit $\sim$66\% \label{mesDistrib}}
     \end{center}
 \end{figure}

The false alarm population is dominated by TCEs with 3 transits and low detection strength (MES). Artifacts with large MES are more effectively identified and removed by TPS.  In Figure~\ref{mesDistrib} we plot the fraction of planet candidates with 3 transits as a function of MES. The majority of 3 transit events have MES$<$8. If we assume for the sake of simplicity that all planet candidates with three transits are false alarms, our reliability at MES$<$8 is 66\%.
This number is a rough estimate, but serves as reminder that, because TCERT errs on the side marking ambiguous cases as candidates, our planet sample is contaminated by design with events not caused by planets.

\subsection{Multiple-KOI Systems}
With this catalog there are several new multi-KOI systems.
Here we give a brief overview and identify differences between this catalog and the catalog of \citet{Burke14}, which used roughly half the data used in this work.

\begin{figure}
    \includegraphics[width=0.45\textwidth]{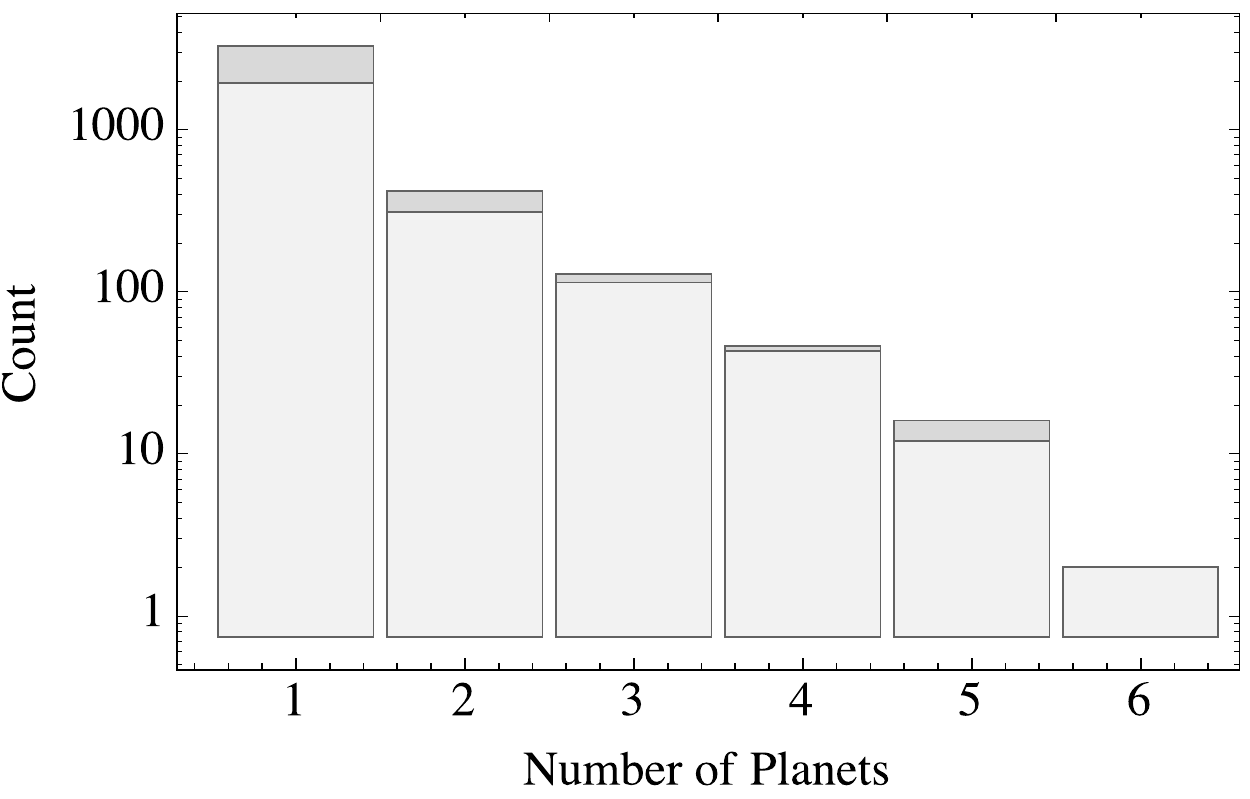}
    \medskip
    \includegraphics[width=0.45\textwidth]{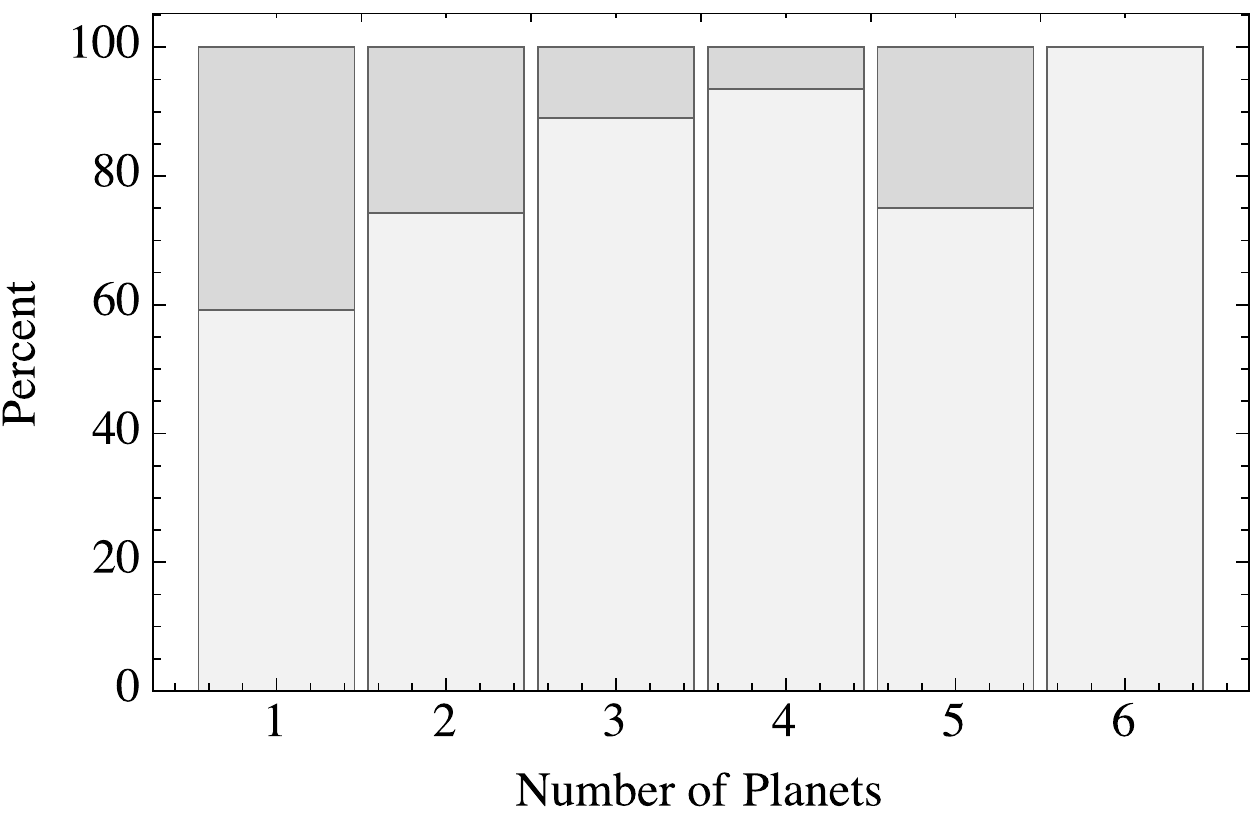}
    \caption{A histogram of the number of KOI systems as a function of multiplicity.  Results from the \citet{Burke14} catalog are shown in light gray while results from this work are darker gray.  Raw counts are shown in the top panel.  The bottom panel shows the relative contributions to the total of the two catalogs.\label{multiHist}}
\end{figure}

For this comparison, we select all multi-KOI systems in each catalog that do not have any planet pairs with a period ratio smaller than 1.1, eliminating putative systems that are likely to be dynamically unstable or split multi-planet systems such as Kepler-132 (K00284).  In the Q8 catalog there are \oldsys\ unique KOI systems where \oldmulti\ are multi-KOI systems, these comprise \oldplans\ total planet candidates with \oldmultiplans\ candidates in multi-KOI systems.  Our catalog increases this yield to \newsys\ total KOI systems with \newmulti\ multi-KOI systems comprising \newplans\ total planet candidates with \newmultiplans\ appering in multi-KOI systems.  The multiplicity of these 128 new multi-KOI systems include a net gain of 107 two-planet, 14 three-planet, 3 four-planet, and 4 five-planet systems.  Figure \ref{multiHist} shows a histogram of the previous and new mult-planet systems as a function of multiplicity.

\subsection{Habitable Zone Planets}

The habitable zone (HZ) planets, those that orbit their parent stars at distances that allow liquid water to exist on their surface, are of particular interest for follow-up, validation, and confirmation. The first theoretical limits on the extent of the HZ come from \citet{Kasting93}. Confirmed small planets in or near the HZ that were  originally found in \kepler\ data include Kepler-69c \citep{Barclay13}, Kepler-62f \citep{Borucki13}, and Kepler-186f \citep{Quintana14}.

To help identify potential habitable zone KOIs, we provide an estimate of equilibrium temperature, \teq, in the tables at the NASA Exoplanet Archive.
We calculate
\begin{equation}
T_{\rm eq} = T_{\rm eff}(R_{*}/2a)^{1/2}[ f(1-A_{\rm B})]^{1/4}
\end{equation}
where $T_{\rm eff}$ and $R_{*}$ are the effective temperature and radius of the parent star, $a$ is the assumed orbital semi-major axis, $f$ parameterizes thermal circulation from the day to night sides, and $A_{\rm B}$ is the Bond albedo. We assume full thermal circulation ($f=1$) and $A_{\rm B} = 0.3$.
\citet{Kopparapu13} suggest using incident stellar insolation relative to the Earth, $S_{\rm eff}$, in preference to \teq. Values of \teq\ reported at the archive can be converted to insolation flux with the formula

\begin{equation}
S_{\rm eff} =   \left( \frac{T_{\rm eq}} {T_{\earth}} \right) ^{4}
\end{equation}
where $T_{\earth}$ is 255\,K, the computed equilibrium temperature of the Earth given our assumptions. \citet{Kopparapu13} predict an inner edge to the HZ in the Solar System of $S_{\rm eff}$ between 1.015 (for a conservative, cloud free model) and 1.77 (based on models of early Venus), corresponding to \teq\ between 255 and 294\,K.

Within the HZ, planets small enough to have a solid surface (rocky planets), are of particular interest. \citet{Rogers14} determined the upper bound for the radius of a rocky planet is 2.0\,\Rearth with 68\% confidence, based on the extensive radial velocity follow-up of small \kepler\ candidates by \citet{Marcy14}.

\begin{figure}[tbh]
     \begin{center}
    \includegraphics[angle=0, scale=.35]{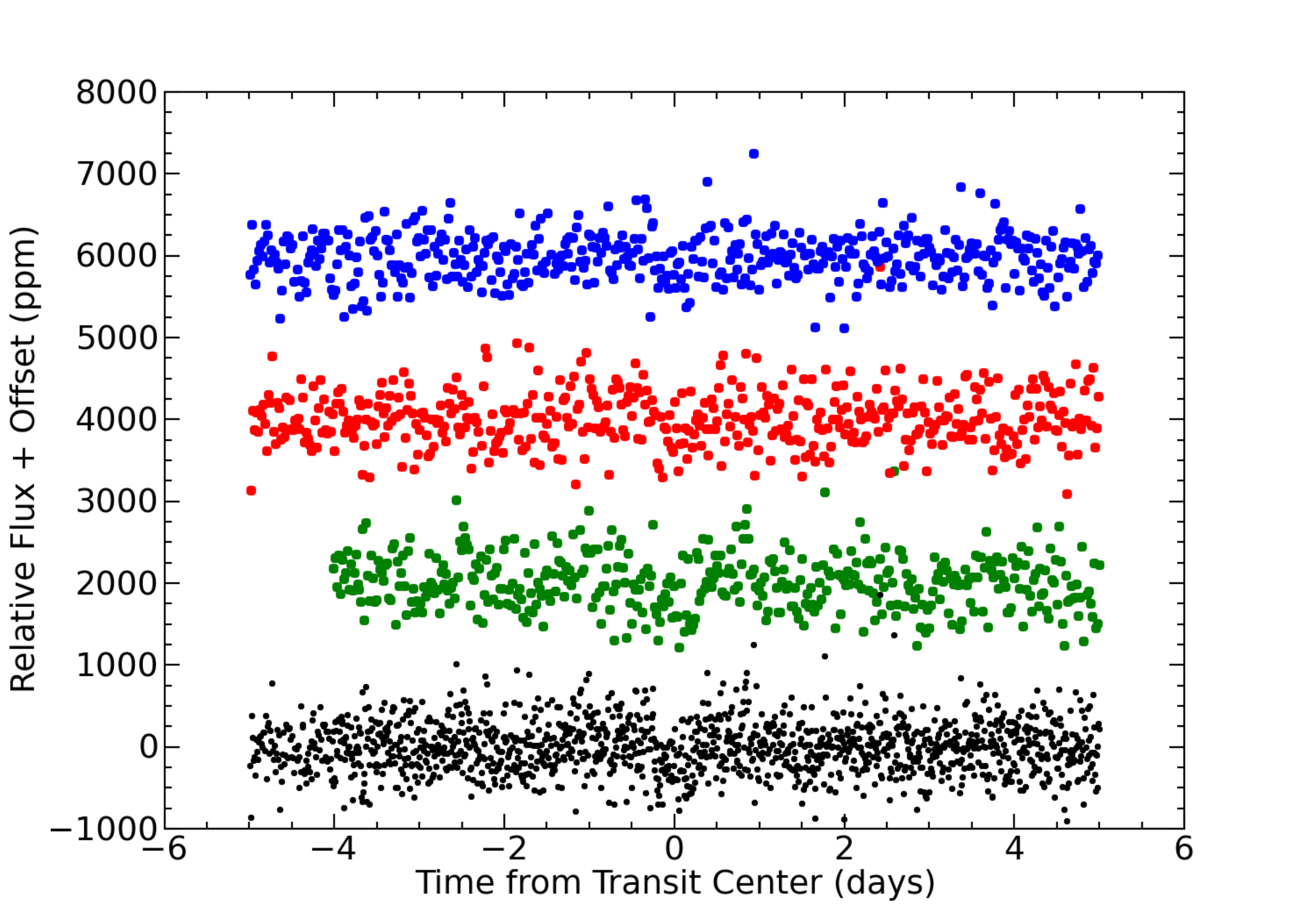}
    \caption{An example KOI at the detection threshold, K05202.01. The bottom section shows the folded lightcurve centered on the transit. The upper sections show the individual
events, vertically offset for clarity. The transit is barely detected, with a MES of 7.8, and the individual events are scarcely visible. Because there is no clear reason to mark this KOI as a false positive it is marked as a planet candidate. \label{lowsnrkoi}}
     \end{center}
 \end{figure}

In this section, we highlight a few individual candidates with $R < 2$\,\Rearth\ and \teq\ that places them in or near the HZ.
K05202.01 ($R = 1.8$\,\Rearth, \teq = 227\,K),  %weak candidate  226.9  3t
K05506.01 ($R = 1.6$\,\Rearth, \teq = 230\,K),  %low snr 229.7         3t
K05856.01 ($R = 1.7$\,\Rearth, \teq = 280\,K),  %low snr 281.1  5   5
K06151.01($R = 1.4$\,\Rearth, \teq = 210\,K)  % dodgy pass 210.7  3
are among our lowest SNR detections. We show an example lightcurve in Figure~\ref{lowsnrkoi}. Although they are detected with sufficient formal significance, it is challenging to distinguish genuine transits from other effects in the lightcurve at this noise level. Less vetting is possible on such shallow KOIs, and the possibility of misidentification as a planet candidate is correspondingly higher.

%Latex table autogenerated by
%$Id: hzTable.py 58190 2015-02-02 23:29:11Z fmullall $
%$URL: svn+ssh://murzim/repo/so/trunk/vetting/q1q16/paper/table/hzTable.py $
\begin{deluxetable}{lrrrrr}
\tablewidth{0pt}
\tabletypesize{\normalsize}
\tablecaption{
Table of Small HZ candidates \label{hzTable}
}
 
\tablehead{
    \colhead{KOI}&
    \colhead{Radius}&
    \colhead{\teq}&
    \colhead{SNR}&
    \colhead{\teff}&
    \colhead{$K_{\rm\-p}$}\\
    \colhead{~}&
    \colhead{($\mathrm{R_{\oplus}}$)}&
    \colhead{(K)}&
    \colhead{~}&
    \colhead{(K)}&
    \colhead{~}
}
\startdata
K02184.02 & 1.89$^{+0.26}_{-0.13}$ & 288$^{+33}_{-16}$ & 9.2 & 4893 & 15.5 \\
K02194.03 & 1.43$^{+0.56}_{-0.12}$ & 239$^{+66}_{-16}$ & 12.1 & 6038 & 13.9 \\
K05068.01 & 1.57$^{+0.52}_{-0.23}$ & 290$^{+70}_{-26}$ & 15.1 & 6440 & 13.1 \\
K05236.01 & 1.98$^{+0.98}_{-0.18}$ & 240$^{+87}_{-16}$ & 22.5 & 6241 & 13.1 \\
K05737.01 & 1.32$^{+0.51}_{-0.14}$ & 254$^{+72}_{-18}$ & 6.3 & 5916 & 13.8 \\
K05805.01 & 1.84$^{+0.24}_{-0.12}$ & 174$^{+21}_{-8}$ & 14.2 & 5192 & 14.5 \\

\enddata
\tablecomments{
{\footnotesize Table of strong candidates for rocky HZ planets. See text for selection criteria, and the Exoplanet Archive for the full set of parameters.}
}
\end{deluxetable}

In Table~\ref{hzTable} we list our strongest candidates for rocky HZ planets.
In addition to our criteria for radius and \teq, these KOIs are either detected with MES $>8$ or with $>3$ transits. They have measured impact parameters $<0.8$ (to guard against unreliable fits), and the transit is visible by eye in the PDC lightcurve. This reduces the risk of artifacts contaminating our sample of strongest candidates. These are our best candidates for habitable, rocky worlds, and prime candidates for confirmation and validation. We note that the estimated \teq\ for K05805.01 is colder than outer limit of the HZ computed by \citet{Kopparapu13} based on evidence that water may have existed on Mars in the past (190\,K given our assumptions).

Two KOIs from Table~\ref{hzTable} are in a multi-KOI systems. \citet{Lissauer12}
argue almost all candidates in multiple systems, such as these, are true planets. The K02184 system contains two {\add KOIs}, one in the HZ and one with a period of 2 days (K02184.01).
This system is the exception to the rule in that K02184.01 is a false positive due to a background event. The source of the K02184.02 event is consistent with being on the target star.
The K02194 system has three planet candidates, all smaller than 2\Rearth. Because the \citet{Lissauer12} analysis does not include a treatment of non-astrophysical false alarms \citep{Rowe14multis}, which is a concern for the outer most candidate, we do not claim K02194.03 has been statistically validated, although it has the strongest claim of any KOI in the table to being due to a transiting planet.

Even if confirmed or validated, further work is required to confirm the status of these KOIs as rocky, habitable zone planets. These KOIs have large ($\sim 25$\%) uncertainties in their stellar radii, which corresponds to a similar uncertainty in planet radius, and a $\sim 30$\% uncertainty in \teq.  With these uncertainties, we can not say that our measurements securely identify any of these KOIs as rocky, HZ candidates. However, given the size of the sample it is likely that at least one (or more) of these KOIs has a true radius and \teq\ that meets our criteria. Follow-up observations are needed to constrain stellar radii and identify the true properties of these candidates.
Until such follow-up work is complete, these KOIs represent the best, closest analogs to the Earth known to date.

\section{Conclusion}

We describe the latest planet candidate catalog for the \kepler\ mission.
With four years of data, \kepler\ is now sensitive to smaller, and longer period planets than before. We discuss some of the caveats that users should be aware of when using the catalog, such as our choice to err on the side of including a KOI as a planet candidate in the face of uncertain evidence, and the challenges of vetting extremely deep and shallow transit events.

The false alarm rate increases for small, long period planets due to additional sources of spurious events in this regime. Although we eliminate the majority of false alarms and false positives, some remain in the final catalog, particularly at low signal to noise.

We highlight a handful of possibly rocky planets in or near the HZs of their parent stars, including K02194.03, the third candidate smaller than 2\,\Rearth\ in a multiple system, and K05805.01 whose \teq\ suggests it is colder than the outer HZ limit proposed by \citet{Kopparapu13}.

We make our first steps towards automating the identification of planet candidates and false positives, which will help remove some of the subjectivity and human error of previous releases. The full table of KOIs is available at the Exoplanet Archive hosted at \nexsci.

\acknowledgements
Funding for this Discovery mission is provided by NASA’s
Science Mission Directorate.
Some of the data presented in this paper were obtained from the Mikulski Archive for Space Telescopes (MAST). STScI is operated by the Association of Universities for Research in Astronomy, Inc., under NASA contract NAS5-26555. Support for MAST for non-HST data is provided by the NASA Office of Space Science via grant NNX13AC07G and by other grants and contracts.
This research has made use of the NASA Exoplanet Archive, which is operated by the California Institute of Technology, under contract with the National Aeronautics and Space Administration under the Exoplanet Exploration Program.
 We would like to thank the Exoplanet Archive staff for their efforts in supporting the Kepler pipeline data products.
 Funding for the Stellar Astrophysics Centre is provided by The Danish National Research Foundation (Grant agreement no.: DNRF106). The research is supported by the ASTERISK project (ASTERoseismic Investigations with SONG and Kepler) funded by the European Research Council (Grant agreement no.: 267864).

\appendix

NASA missions tend to accumulate acronyms. Here we provide a summary of the more important ones used in this paper for easy reference.

\noindent
CAL: Module of the \kepler\ pipeline that calibrates the recorded pixel values.\\
DV: Data Validation. Module of the \kepler\ pipeline that fits the transit parameters and provides diagnostics used in triage and vetting.\\
EB: Eclipsing Binary star.\\
HZ: Habitable Zone. Region around a star where a planet could have surface temperatures consistent with liquid water.\\
KIC: Kepler Input Catalog. This is the catalog used in target selection, and provides stellar parameters for much of the sample.\\
KOI: \kepler\ Object of Interest. A unique identifier of a transit event. Some KOIs are marked as false positive to indicate that the transit event is not due to a planet.\\
MES: Multiple Event Statistic. The signal-to-noise ratio of the detection of a TCE by the TPS module of the pipeline.\\
PA: Photometric Aperture. The pipeline module that extracts photometry.\\
PC:  Planet Candidate.\\
PDC: Presearch Data Conditioning. The pipeline module that removes instrumental signals from lightcurves.\\
TCE: Threshold Crossing Event. A period set of dips in a lightcurve that may be due to a transit.\\
TCERT:  TCE Review Team. A committee that reviews TCEs and KOIs to identify false alarms and false positives.\\
TPS: Transit Planet Search. The module of the pipeline that searches for transits. Potential transits (TCEs) are passed to DV to be fit.\\
TTV: Transit Timing Variations. Irregularly spaced transits due to gravitational interaction in multi-planet systems.\\

{\it Facilities}: \kepler

\bibliographystyle{apj}
\bibliography{q1q16}

\clearpage
%Latex table autogenerated by
%$Id: makeTable.py 57798 2014-12-13 00:20:51Z fmullall $
%$URL: svn+ssh://murzim/repo/so/trunk/vetting/q1q16/paper/table/makeTable.py $
\begin{deluxetable}{lrrrrccccc}
\tablewidth{0pt}
\tabletypesize{\normalsize}
\tablecaption{New KOIs discovered in 16 Quarters of Data}
 
\tablehead{
    \colhead{KOI}&
    \colhead{KeplerId}&
    \colhead{TCE}&
    \colhead{Status}&
    \colhead{V-Shape}&
    \colhead{Flag}&
    \colhead{Period}&
    \colhead{Epoch}&
    \colhead{Depth}&
    \colhead{Radius}\\
    \colhead{~}&
    \colhead{~}&
    \colhead{~}&
    \colhead{~}&
    \colhead{~}&
    \colhead{~}&
    \colhead{(Days)}&
    \colhead{(BKJD)}&
    \colhead{(ppm)}&
    \colhead{($\mathrm{R_{\oplus}}$)}
}
\startdata
\object{K00099.02} & 8505215 & 1 & FP & - & 0 & 79.878049 & 184.770441 & - & -\\
\object{K00129.02} & 11974540 & 2 & FP & - & 0 & 143.211094 & 221.755647 & - & -\\
\object{K00238.03} & 7219825 & 3 & FP & 0.598 & 0 & 362.997(26) & 256.446(45) & 282(31) & $1.86^{+0.31}_{-0.17}$\\
\object{K00266.02} & 7375348 & 2 & PC & 0.666 & 0 & 47.74360(28) & 160.1645(51) &  102.7(6.4) & $1.82^{+0.80}_{-0.63}$\\
\object{K00337.02} & 10545066 & 2 & PC & 0.895 & 0 & 154.6074(43) & 250.520(21) & 280(23) & $1.56^{+0.64}_{-0.13}$\\
\object{K00353.03} & 11566064 & 3 & PC & 0.790 & 0 & 11.16223(13) & 133.523(11) &  98.1(9.3) & $1.38^{+0.86}_{-0.21}$\\
\object{K00365.02} & 11623629 & 2 & FP & 0.743 & 0 & 117.7610(22) & 175.560(16) &  53.9(8.2) & $0.62^{+0.10}_{-0.03}$\\
\object{K00423.02} & 9478990 & 2 & FP & -143.711 & 0 & 360.4776(99) & 253.911(20) & 604(117) & $11943.77^{+2456.24}_{-2393.47}$\\
\object{K00492.02} & 3559935 & 2 & PC & 0.645 & 1 & 265.296(15) & 297.870(24) & 608(134) & $2.99^{+1.16}_{-0.60}$\\
\object{K00520.04} & 8037145 & 4 & PC & 0.530 & 0 & 51.16579(64) & 172.308(11) & 305(25) & $1.54^{+0.55}_{-0.16}$\\

\enddata
\tablecomments{  All new KOIs discovered in \qRange\ data not found
in the earlier \kepler\ catalogs. The full set of fitted parameters
can be found at the Exoplanet Archive (\S~\ref{NEXSCI}); we
show only a summary set here. KOI is a unique identifier assigned
to every \kepler\ Object of interest. Kepler Id is a unique
identifier assigned to every target star in the KIC \citep{Brown11}.
The TCE number indicates the order in which the pipeline found this event
around this target.
A status of FP (false positive) indicates that we believe the KOI is
not a bona-fide planet; PC (Planet Candidate) indicates that we have
no compelling evidence that the signal is not due to a planet.
The v-shape statistic is
not included at the Exoplanet Archive and is described
in \S~\ref{CAVEATS}. The flag column is set for KOIs detected with three transits
and MES $<$8. As discussed in \S~\ref{human}, we expect lower reliability for
this population of candidates.
The numbers in parentheses indicate the 1\,$\sigma$
uncertainty in the two least significant digits. For example,
1.23(45) = 1.23 $\pm$ 0.45.
In cases where the MCMC fit failed to converge we
report the period and epoch of the detection only. Some KOIs,
such as K00423.02, have extremely large reported radii.
These are either extremely v-shaped transits
that are difficult to fit with a planet model, or low SNR false alarms
where the MCMC fits struggle to converge on a sensible value.
(This table in available in its entirety in a machine-readable form in the online journal.
A portion is shown here for guidance regarding its form and content.)
\label{newKoiTable}
%Created with $Id: makeTable.py 57798 2014-12-13 00:20:51Z fmullall $
%$URL: svn+ssh://murzim/repo/so/trunk/vetting/q1q16/paper/table/makeTable.py $
}
\end{deluxetable}
\clearpage

\clearpage
%Latex table autogenerated by
%$Id: makeTable.py 57798 2014-12-13 00:20:51Z fmullall $
%$URL: svn+ssh://murzim/repo/so/trunk/vetting/q1q16/paper/table/makeTable.py $
\begin{deluxetable}{lrrrrccccc}
\tablewidth{0pt}
\tabletypesize{\normalsize}
\tablecaption{Previously Known KOIs vetted with 16 Quarters of Data}
 
\tablehead{
    \colhead{KOI}&
    \colhead{KeplerId}&
    \colhead{TCE}&
    \colhead{Status}&
    \colhead{V-Shape}&
    \colhead{Flag}&
    \colhead{Period}&
    \colhead{Epoch}&
    \colhead{Depth}&
    \colhead{Radius}\\
    \colhead{~}&
    \colhead{~}&
    \colhead{~}&
    \colhead{~}&
    \colhead{~}&
    \colhead{~}&
    \colhead{(Days)}&
    \colhead{(BKJD)}&
    \colhead{(ppm)}&
    \colhead{($\mathrm{R_{\oplus}}$)}
}
\startdata
\object{K00157.01} & 6541920 & 2 & PC & 0.933 & 0 & 13.024928(13) & 138.17627(83) &  809.2(6.8) & $2.94^{+0.21}_{-0.18}$\\
\object{K00157.02} & 6541920 & 3 & PC & 0.956 & 0 & 22.687141(25) & 148.45544(92) &  995.5(8.0) & $3.26^{+0.24}_{-0.20}$\\
\object{K00157.03} & 6541920 & 1 & PC & 0.854 & 0 & 31.995506(28) & 154.16149(72) & 1411(12) & $3.89^{+0.28}_{-0.24}$\\
\object{K00157.04} & 6541920 & 5 & PC & 0.277 & 0 & 46.68580(13) & 225.0421(20) & 605(11) & $2.74^{+0.20}_{-0.17}$\\
\object{K00157.05} & 6541920 & 4 & PC & 0.861 & 0 & 118.37838(31) & 287.2891(20) & 1117(14) & $3.46^{+0.25}_{-0.22}$\\
\object{K00157.06} & 6541920 & 6 & PC & 0.972 & 0 & 10.304005(17) & 138.5042(14) &  320.2(6.1) & $1.85^{+0.13}_{-0.12}$\\
\object{K00298.02} & 12785320 & 2 & PC & 0.180 & 0 & 57.38397(29) & 170.7285(43) & 237(13) & $1.68^{+0.00}_{0.00}$\\
\object{K00536.01} & 10965008 & 1 & PC & 0.418 & 0 & 81.16943(32) & 178.5932(31) & 1233(26) & $3.41^{+1.37}_{-0.29}$\\
\object{K00638.01} & 5113822 & 1 & PC & 0.863 & 0 & 23.642191(25) & 172.58943(85) & 1171(11) & $3.87^{+1.49}_{-0.61}$\\
\object{K00638.02} & 5113822 & 2 & PC & 0.928 & 0 & 67.09330(14) & 146.5670(15) & 1294(16) & $4.06^{+1.57}_{-0.63}$\\

\enddata
\tablecomments{  Previously known KOIs revetted with 16 quarters of data. A
known KOI was revetted if it, or any other KOI around the same target star
has a period $>$ 50\,days. Some additional KOIs also received further scrutiny, and their new
statuses are included here.
See Table~\ref{newKoiTable} for a description of the columns.
(This table in available in its entirety in a machine-readable form in the online journal.
A portion is shown here for guidance regarding its form and content.)
\label{cjbTable}
%Created with $Id: makeTable.py 57798 2014-12-13 00:20:51Z fmullall $
%$URL: svn+ssh://murzim/repo/so/trunk/vetting/q1q16/paper/table/makeTable.py $
    }
\end{deluxetable}
\clearpage

\end{document}